# Towards advancing the earthquake forecasting by machine learning of satellite data

Pan Xiong [1, 8], Lei Tong [3], Kun Zhang [9], Xuhui Shen [2, *], Roberto Battiston [5, 6], Dimitar Ouzounov [7], Roberto Iuppa [5, 6], Danny Crookes [8], Cheng Long [4] and Huiyu Zhou [3]

[1] Institute of Earthquake Forecasting, China Earthquake Administration, Beijing, China

[2] National Institute of Natural Hazards, Ministry of Emergency Management of China, Beijing, China

[3] School of Informatics, University of Leicester, Leicester, United Kingdom

[4] School of Computer Science and Engineering, Nanyang Technological University, Singapore

[5] Department of Physics, University of Trento, Trento, Italy

[6] National Institute for Nuclear Physics, the Trento Institute for Fundamental Physics and Applications, Trento, Italy

[7] Center of Excellence in Earth Systems Modeling & Observations, Chapman University, Orange, California, USA

[8] School of Electronics, Electrical Engineering and Computer Science, Queen's University Belfast, Belfast, United Kingdom

[9] School of Electrical Engineering, Nantong University, Nantong, China

\*   Correspondence: Xuhui Shen (shenxh@seis.ac.cn)

**Highlights**

- An AdaBoost-based ensemble framework is proposed to forecast earthquake

- Infrared and hyperspectral global data between 2006 and 2013 are investigated

- The framework shows a strong capability in improving earthquake forecasting

- Our framework outperforms all the six selected baselines on the benchmarking datasets

- Our results support a Lithosphere-Atmosphere-Ionosphere Coupling during earthquakes

**Abstract**

Earthquakes have become one of the leading causes of death from natural hazards in the last fifty years. Continuous efforts have been made to understand the physical characteristics of earthquakes and the interaction between the physical hazards and the environments so that appropriate warnings may be generated before earthquakes strike. However, earthquake forecasting is not trivial at all. Reliable forecastings should include the analysis and the signals indicating the coming of a significant quake. Unfortunately, these signals are rarely evident before earthquakes occur, and therefore it is challenging to detect such precursors in seismic analysis. Amongst the available technologies for earthquake research, remote sensing has been commonly used due to its unique features such as fast imaging and wide image-acquisition range. Nevertheless, early studies on pre-earthquake and remote-sensing anomalies are mostly oriented towards anomaly identification and analysis of a single physical parameter. Many analyses are based on singular events, which provide a lack of understanding of this complex natural phenomenon because usually, the earthquake signals are hidden in the environmental noise. The universality of such analysis still is not being demonstrated on a worldwide scale. In this paper, we investigate physical and dynamic changes of seismic data and thereby develop a novel machine learning method, namely Inverse Boosting Pruning Trees (IBPT), to issue short-term forecast based on the satellite data of 1,371 earthquakes of magnitude six or above due to their impact on the environment. We have analyzed and compared our proposed framework against several states of the art machine learning methods using ten different infrared and hyperspectral measurements collected between 2006 and 2013. Our proposed method outperforms all the six selected baselines and shows a strong capability in improving the likelihood of earthquake forecasting across different earthquake databases.

**Keywords**

Earthquake forecasting; Earthquake precursors; Machine learning; Infrared and hyperspectral parameters

## 1. Introduction

Over 500,000 earthquakes are recorded per year. Many of these are undetected or unnoticed because of their small magnitude, while others cause devastation to buildings, bridges and mountains. Earthquake damage mitigation is an active topic in geological research. The majority of earthquakes occur due to the sudden release of stress in the earth's crust that gradually builds up from tectonic movement. However, the response of the crust to the changing stress is non-linear and is dependent on the compression capability of the crust, which is highly variable and complex (Council 2003). Satellite remote sensing enables us to detect large-range and continuous changes of the near-surface thermal field (Frick and Tervooren 2019; Niu et al. 2012; Ouzounov et al. 2006; Pulinets et al. 2006; Tramutoli et al. 2005; Tronin 2007). It can be utilized to monitor thermal anomalies caused in the process of earthquake preparation, which provides useful hints learnt from the measurements for short term and imminent forecasting of earthquakes. Moreover, some researchers believed that thermal anomalies might be related to the variation in the composition of atmospheric gas mixtures above the near-surface fault (Pulinets and Ouzounov 2018). Micro-fractures of rock and surface within earthquake regions may expand due to the intensification of pre-earthquake tectonic activities, which helps release underground high-concentration gas, such as $H_2$, CO, $O_3$, $CO_2$ and water vapour to space, and produce additional atmospheric electricity, which stimulates the infrared electromagnetic radiation (Liperovsky et al. 2011).

Existing studies mainly aim at abnormality identification and analysis of a single physical parameter or a specific earthquake; the results of analyzing abnormalities are short of universal, and cannot confidently explain the pre-earthquake multi-parameter anomalies. In recent years, in order to study the pre-earthquake multi-parameter anomalies from the perspective of the energy balance of the Earth, Pulinets and Ouzounov have revised the physical model of lithosphere-atmosphere-Ionosphere-coupling (LAIC) (Ouzounov et al. 2018a; Pulinets and Ouzounov 2011; Wu et al. 2012) which explains the synergy of different physical processes and variations, known as short-term pre-earthquake anomalies. Other scholars modify

LAIC into lithosphere-coversphere-atmosphere (LCA) concept supporting a wide range of remote sensing applications (Wu et al. 2012). It has been pointed out that the effect of seismic crustal stress enables mechanical energy to be transformed to heat energy, and then transmitted to the earth surface through pores in rocks and tiny pre-earthquake ruptures, resulting in enhancement of stress in the rock mass and increase of surface temperature in strike-slip parts. At the same time, the local stress enhancement and the occurrence of micro-earthquakes can cause numerous micro-cracks and micro-fractures inside the lithosphere, so that the geo- gases and fluids within the lithosphere, such as $H_2$, Rn, CO, $O_3$, $CO_2$, and water vapour, would spill out along these channels. The increase of geo-gases in the atmosphere stimulates physical processes and chemical reactions from the ground surface up to the troposphere. Along with the effect of electric fields, it stimulates the infrared electromagnetic radiation, and the atmospheric temperature may increase, which can be measured by satellites (Pulinets and Ouzounov 2018).

With in-depth studies on the LCA coupling model as well as accumulation and utilization of pre-earthquake satellite multi-parameters (Ouzounov et al. 2018a; Wu et al. 2012), some correlation analysis of multi-parameters has been carried out (Table S1). Singh et al. analyzed abnormal variations of sea surface temperature (SST), surface latent heat flux (SLHF), atmospheric temperature and humidity before the earthquake occurred in Sumatra, Indonesia on December 26, 2004 (Singh et al. 2007) and the Wenchuan earthquake on May 12, 2008 (Singh et al. 2010); it was considered that there was a strong land-ocean-atmosphere coupling before earthquakes. Rawat et al. compared anomalies of temperature and long-wave radiation before the earthquakes happened in India and Romania (Rawat et al. 2011). Wu et al. summarized the studies of pre-earthquake anomalies before the earthquake hit L'Aquila, Italy on 22 October 2012, where they selected the parameters of the lithosphere, coversphere, atmospheric and ionized layers to study abnormal synchronization and coupling mechanisms (Wu et al. 2016). Jingfeng et al. studied the abnormal synchronization of multi-parameters such as the latent heat flux, long-wave radiation, atmospheric temperature, humidity and pressure before the Wenchuan Earthquake in 2008 (Jing et al. 2013). Qin et al.

utilized multiple satellite parameters to analyze thermal anomalies before the New Zealand Earthquake in 2010-2011, and they studied the physical mechanism behind the earthquake thermal anomaly by investigating the regional tectonics, hydrogeology and meteorological environment (Qin et al. 2012). They also proposed the deviation-timespace-thermal method to analyze the spatiotemporal synchronicity and inherent mechanism among multi-parameters of various thermal anomalies and ionospheric anomalies before the 6.7 magnitude's earthquake occurred in Pu'er (Qin et al. 2013b) and Yushu (Qin et al. 2013a). Ouzounov et al. tested (retrospectively and prospectively) a new approach of integrated satellite and terrestrial framework (ISTF) for detecting atmospheric pre-earthquake signals. The approach is based on a sensor web of coordinated analysis between three physical parameters validated by the LAIC model: OLR (satellite), dTEC (electron concentration in the ionosphere), and atmospheric chemical potential (atmospheric assimilation models). ISTF has been applied for three major earthquakes: M 6.0 Napa of 2014 (USA), M 6.0 Taiwan of 2016, and M 7.0 Kumamoto, Japan of 2016. Molchan's error diagram (MED) for all parameters shows results that are better than random guesses. Prospective tests based on 22 earthquakes over Japan (2014-15) revealed the existence of general temporal-spatial evolution pattern in the atmosphere ahead of the main earthquakes only in cases when a multi-parameter analysis been used (Ouzounov et al. 2018b).

However, the above studies have two several constraints such as (1) earthquake studies are still limited in space and time, and (2) the methodology for multi parameters analysis is non uniform and therefore cannot meet the requirements of the practice of earthquake monitoring. Moreover, the applications of anomaly-evaluation applied to specific earthquakes often are with lack of understanding of the underlying physics, which may cause various or even contradictory conclusions for the same earthquake (Blackett et al. 2011a, b).

With the rapid development of artificial intelligence and machine learning, researchers have made progress in the domain of earth sciences (Sarkar and Mishra 2018), especially in earthquake forecasting (Asencio–Cortés et al. 2018; Asim et al. 2018a, b; Bergen et al. 2019; Hulbert et al. 2018; Lubbers et al. 2018; Rafiei and Adeli 2017; Reyes et al. 2013; Rouet-Leduc et al. 2018). At the same time, machine learning are also very effective for spatial remote sensing data handling (Du et al. 2020). For example, artificial intelligence technology may provide a useful measure for resolving those problems mentioned above. In this paper, a novel Inverse Boosting Pruning Trees (IBPT) based framework is presented for earthquake forecasting, which utilizes satellite data with ten parameters such as infrared sensing, hyperspectral imaging and gas sensing signals collected from worldwide earthquakes during 2006 and 2013. Our proposed method aims to be a general-purpose technology, using the labels generated by time series clustering techniques. This technology simplifies the process of forecasting because the proposed method deals with the sequences of labels instead of real data itself. Four datasets of earthquakes with different magnitudes, collected between 2006 and 2013, are used to forecast earthquakes and compared against other state of the art techniques.

**2. Data and processing**

*2.1 Datasets*

Considering the research results reported in the literature (Jing et al. 2013; Qin et al. 2013a; Qin et al. 2012; Qin et al. 2013b; Rawat et al. 2011; Singh et al. 2007; Singh et al. 2010; Wu et al. 2016) and published data, without loss of generality, according to the lithosphere-coversphere-atmosphere (LCA) coupling model (Ouzounov et al. 2018b; Pulinets and Ouzounov 2011; Wu et al. 2012), we selected ten parameters for the earthquake anomaly analysis. Figure S1 shows the inherent relations between the selected ten parameters. These parameters were generated from two different satellite data sources (Table S2). The first nine parameters as shown in Table S2 were created from the Atmospheric Infrared Sounder (AIRS) on NASA's spacecraft Aqua and the engaged parameters are recorded with 1.0 × 1.0° resolution, with the frequency of twice per day.

Specifically, surface skin temperature, temperature of the atmosphere at the earth's surface, water vapour mass mixing ratio at the surface, total integrated column ozone burden, retrieved total column co, retrieved total column $CH_4$, ARIS outgoing longwave radiation flux, and clear-sky outgoing longwave radiation flux can be obtained from the AIRS3STD v6 product (L3 Standard Daily Product processed using only AIRS radiances in Version 6), and land surface temperatures is from the AIRX3SPD v6 product (L3 Support Daily Product processed using AIRS and AMSU radiances in Version 6), which retrieved from MODIS averaged over MYD11C3 (MODIS/Aqua Land-Surface Temperature/Emissivity Monthly Global 0.05Deg CMG) 0.05 degree (~5 km) pixels. The last parameter was obtained from the National Oceanic and Atmospheric Administration (NOAA) Climate Forecasting Center web site (ftp ftp.cpc.ncep.noaa.gov ; cd precip/noaa* for OLR directories), which provided original gridded daily Outgoing Longwave Radiation (OLR) data from NCAR with temporal interpolation. The OLR algorithm for analyzing the Advanced Very High-Resolution Radiometer (AVHRR) data is proposed by Gruber and Krueger (Gruber and Krueger 1984), which integrates the IR data with the wavelengths between 10 and 13 μm. The data is mainly sensitive to the near surface and/or cloud temperatures. The two data sources provide abundant observation data, and all the ten parameters are of the same spatial resolution and time scales. This study covers worldwide events, including land and submarine earthquakes. In total, 1,234 earthquakes with magnitudes between 6 and 7, and 137 earthquakes with magnitude 7 and over, are recorded in the study area, which is spread over the period between 2006 and 2013. In order to verify the reliability and improve the robustness of the proposed model, we generate 1,371 artificial non-earthquake events, the same amount as the real earthquakes, and stagger the time and place to match when and where the real earthquakes occur. As can be seen, there are millions of measurement values in this study.

We look at the temporal features that are extracted within $N$ days before an earthquake occurs, and attempt to detect earthquake anomalies during these days. As reported by Tronin et al., anomalies are observed around 6 ~ 24 days before an earthquake strikes (Gorny et al. 2020). Given that there is no universal standard for the temporal window, we set the temporal window to be 30 by default in our study and determine the best temporal window in the experiments. The spatial feature is $M$ degrees away from the epicenter where the earthquake occurs. By analyzing the NOAA and Moderate Resolution Imaging Spectroradiometer (MODIS) images before earthquakes, Ouzounov et al. found that thermal anomalies occur approximately 2.5° away from the epicenter (Ouzounov et al. 2007). Again, since there is not any agreed standard, the square region with its center at the epicenter and a deviation of 3° was selected as the spatial feature in our study.

*2.2. Features generation*

The original satellite data is in a format that is not appropriate for the proposed algorithm to proceed and requires data preprocessing. The first task to be completed is to split the dataset: each dataset is carefully split into two contiguous pieces: 80% for training models, which is 01 January 2006 to 12 May 2012, and 20% for testing purposes and final evaluation, which is from May 12, 2012, until December 25, 2013. The next step is the normalization of data, which is used for the clustering process, the normalization of the datasets needs to be conducted to reduce or eliminate data redundancy, Z-Score normalization was performed on the training data, and normalizing the test data with the normalizing parameters used for training data. Moreover, satellite data is affected by factors such as satellite payload interference and space environment, which can cause occasional errors in continuous data. In our study, we use a "sliding time window" implementation combined with time series clustering techniques (Petitjean et al. 2011) to take overlapping 5-days windows in the time series (separated by a 1-day time lag) features, which ensures that there is a coincidence between the formed series, but also avoids the error of continuous single-point data and has stable robustness.

2.2.1 Standard features' generation

For the purpose of comparison, we generate standard features. Firstly, we choose the well-known scalable K-means algorithm (Lloyd 1982) to classify the data set, the K-means algorithm requires that the user provides the number of clusters to be created. However, this number is a priori unknown, and its selection and evaluations on the results obtained by clustering are crucial. Thus, the most challenging problem of the clustering realm is to select the right number of clusters of the dataset. For these reasons, the Elbow method (Ketchen and Shook 1996) has been applied to the data in order to determine how many groups the original continuous dataset has to be split into.

Secondly, we perform k-means clustering for different values of k, for instance, by varying k from 5 to 20 clusters. For each k, we calculate the total within-cluster sum of square (wss). Further, an algorithm has been implemented and applied to estimate the location of a bend (knee) within all the calculated wss with the number of clusters k.

Finally, clustering labels with the standard features (Figure S2) are generated after the above data processing, which will be used as input in the next step.

2.2.2. Time series based features generation

2.2.2.1. Sliding time window implementation

Figure S3 shows the principle of the sliding time window. The key variables involve the size as well as the sliding step-length of the time windows for data partitioning. The size configuration of the time windows has an essential impact on computational efficiency. In addition, the sliding step also has a critical influence on the clustering results. Too small steps may lead to redundant repetitive computation because of the overlapping of cross-window time series data sets. In our study, the size of the windows is set to 5-days and the sliding step is set to 4-days, which will form overlapping 5-days windows in the time series (separated by a 1-day time lag).

2.2.2.2. Clustering Technique

In order to reduce the data complexity of the series data, we cluster each parameter into several disjoint intervals. Clustering is a difficult task due to the great number of possible geometric shapes for the clusters and distances that can be divided.

For the series-based data, we use classical Dynamic Time Warping (DTW) time series clustering algorithm (Petitjean et al. 2011) to carry out clustering analysis to obtain the demarcation point of each segmentation interval. As the parameters of the sliding time window are configured, the window data of the time series will be extracted and stored for each remote sensing parameter. Then, for each collection of remote sensing time series data, all-time series within the same time window will go through clustering, and each process of clustering within the time window will produce several clusters. As a result, the method will generate a number of cluster labels that cover all the time series data.

However, this number of clusters is a priori unknown, and its selection and later evaluations of the results obtained by the clustering are crucial. Thus, one of the most challenging problems in the clustering realm is to select the right number of clusters for the data sets. For these reasons, the Davies-Bouldin Index (Davies and Bouldin 1979) has been applied to the data in order to determine the optimal number of clusters by varying the number of clusters k from 10 to 20 clusters. For each k, we calculate and compare the corresponding Davies-Bouldin Indexes. While the Davies-Bouldin index reaches its minimum value, the corresponding clusters number k is generally considered as the number of the clusters (Table S3).

Clustering labels with 5 days as the overlapping time series are generated after the above data processing, which will be feed to machine learning algorithms as input features in the next step.

After the data preprocessing is completed, we get four base data sets (Table S4), which are DataSet I: Satellite data of earthquakes of magnitude 7 or greater, DataSet II: Satellite data of earthquakes of magnitude between 6 and 7, DataSet III: the satellite data of the earthquakes of magnitude 7 or greater with the standard features and DataSet IV: the satellite data of earthquakes of magnitudes between 6 and 7 with the standard features. Moreover, we generate two datasets DataSet I-nonoverlap (Dataset II-nonoverlap) exactly as we generate DataSet I (DataSet II) except that we use non-overlapping sliding windows instead of overlapping ones.

**3. Methodology**

The methodology carried out in this work is shown in a schematic way in Figure S4. First, a total of 1234 earthquakes with magnitude between 6 and 7, and 137 earthquakes with magnitude 7 and over, covering a global area, are selected for the study. With a combination of different magnitudes of earthquakes and features, two datasets with ten remote sensing multi-parameters are generated.

Each dataset, is carefully split into the training and test data, and the Z-Score normalization was performed as data preprocessing. Then the "sliding window" technique was implemented for the clustering process. The last step of data preprocessing is clustering, which is first performed on training data, and then cluster the testing sets according to the rules of training data, and finally, we generate time series based features.

We benchmarked eight state of the art methods: Frequent Pattern Learning (FPL) (Cheng et al. 2007), Generalized Linear Models (GLM) (Zeger and Karim 1991), Gradient Boosting Machines (GBM) (Friedman 2001), Deep Neural Network (DNN) (LeCun et al. 2015), Random Forests (RF) (Geurts et al. 2006), Convolutional Neural Network (CNN) (Krizhevsky et al. 2012), Logistic Regression (LR) (Walker and Duncan 1967) and Naive Bayes (NB) (Maron 1961). In our system, which applies Convolutional Neural Networks, we used a network architecture similar to Thibaut Perol's work (Perol et al. 2018) where 4 layers are used in the study as the input samples are too few; this is implemented in Python (v 3.5) with PyTorch (v 0.4.0), and the other eight methods are implemented in R (v 3.4.1) packages: stats, H2O (v 3.18.0.1), arules (v1.6-1) and RevoScaleR (v9.2.1). Since the methods are sensitive to parameter selection, we choose to use the parameters that enable us to obtain the best performance in the experiments. After we have determined the parameters for each method, the performance of each method based on these parameters was compared with the others. We use eight performance measures to evaluate the performance of each method. Benchmarking was performed on a desktop PC equipped with an Intel® Core™ i5-3470 CPU and 16GB of memory.

Each algorithm is trained through the training dataset to produce a parameterized model, which is then applied to the testing dataset for forecasting labels. The models were then applied to every data in each testing dataset, resulting in forecasting label (votes). For the earthquake forecasting, we use majority voting, which is a reasonable decision rule that treats each alternative equally according to May's theorem (May 1952), and every element makes a forecasting vote for the input data and the final earthquake forecasting is the one that receives more than half of the total votes. All the earthquake forecastings are compared with their corresponding actual values. This may result in certain deviations. Such deviations are evaluated resulting in ROC curves and Area Under the Curve (AUC).

*3.1 The proposed machine learning algorithm*

The proposed ensemble model is called Inverse Boosting Pruning Trees (IBPT) scheme, which combines an Adaboost variant with pruning decision trees for classification. In this paper, given the flexibility and ease of use of decision tree, we decide to use decision tree as the boosting base estimator. When an entire tree has a high variance, a decision dump often presents a mismatch problem. Therefore, in order to improve the generalization ability of the model, we consider pruning the tree. Our approach consists of two components: (1) Searching for the best-pruned tree. We applied all the training samples, allowed the decision tree to grow fully, and some branches of the tree are then pruned according to the cost-complexity pruning method mentioned in Breiman (2017). (2) Building an inverse boosting structure. We use an inverse boosting structure with the pruned trees and updated weights. Then, repeat the steps until the maximum number of trees is reached. The proposed framework is summarised in Figure 1

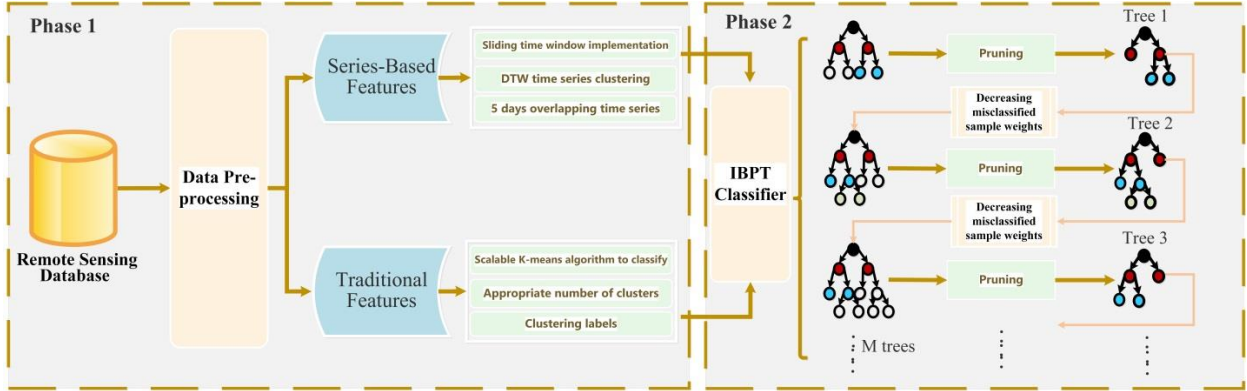

Figure 1 The flowchart of the proposed IBPT framework.

3.1.1. Discrete Adaboost

In this paper, the classification algorithm is based on the discrete Adaboost algorithm proposed by Freund and Schapire (1996). Algorithm 1.1 proposes a baseline scheme of discrete Adaboost, which combines many simple assumptions (called weak learners) to form a strong classifier for classification tasks (Leshem 2005). We summarize the algorithm as follows: (1) Train multiple base classifiers in turn, and distribute the weight $ln(\beta_m)$ according to their training error $\varepsilon_m$. (2) A higher weight $w_{m+1,i}$ is assigned to the samples misclassified by the previous classifier, which will make the classifier pay more attention to these samples. (3) At last, all the weak classifiers and their weights are integrated to constitute an ensemble Learner $G(X)$. Normally, Adaboost uses a decision dump (a one-level decision tree) as its weak learner. But, due to its simple structure, decision dumps sometimes do not fit training data well, result in that the integrated boosting learner does not perform well in complex data sets (Leshem 2005). In this paper, IBPT algorithm is recommended, which outperforms the standard Adaboost algorithm in two aspects: (1) it improves the ability of base classifier fitting and generalization.(2) We propose a new boosting structure to reduce the impact of less contributed data.

---

**Algorithm 1** Discrete Adaboost algorithm.

**Require:** Tree number $M$, $N$ samples

1: Initialise sample weight distribution $D_m = (w_{mi}\ldots), m=1,2,\ldots M, i=1,2,\ldots N,$ and set each sample weight $w_{mi}$ to $\frac{1}{N}$.

2: **for** $m \in (1, M)$ **do**

3:    Fit a classifier $G_m(X)$ to the training data with $D_m$.

4:    Let $d_i = 1$ if the i-th case is classified incorrectly, otherwise zero. Then compute training error $\varepsilon_m = \sum_{i=1}^{N} w_{m,i} d_i$.

5:    Update sample weight for step $w_{m+1,i} = \dfrac{w_{m,i} \beta_m^{d_i}}{\sum_{i=1}^{N} w_{m,i} \beta_m^{d_i}}$, where $\beta_m = (1 - \varepsilon_m)/\varepsilon_m$.

6: **end for**

7: **Output** $G(X) = sign\left(\sum_{m=1}^{M} ln(\beta_m) G_m(X)\right)$.

---

3.1.2. Inverse boosting pruning trees

In this part, we introduce the IBPT algorithm. When an entire tree has a high variance, decision dump often has a high bias against the training data. Therefore, in order to make the system generalized we decide to trim the tree. In our algorithm, we first use all the training samples and allow the decision tree to grow fully, then pry some branches of the tree using the cost-complexity pruning method mentioned in Breiman (2017), and then use the corrected criteria to evaluate the system performance of the pruned tree and the updated weights. Finally, iterate through the steps until reaching the maximum number of trees. To shape our algorithm, here we declare the symbols used in this formula. Here, we represent the training dataset as $L = X_1, y_1, X_2, y_2, \ldots X_N, y_N$, where, $X_n$ means sample feature vector, $y_n$ means class label and $N$ means sample number. We use $D_m = (w_{m,i}, w_{m,i+1}\ldots), m = 1,2, \ldots M, i = 1,2,\ldots,N$ to show the sample weight's distribution in each iteration. $M$ means estimator number (iterations), and in the first iteration of normalization, each sample weight is initialized to $\frac{1}{N}$. In addition, we apply $\varphi_m$ and $G_{final}(x)$ to represent $m$-th estimator's weight and the final classifier, respectively.

3.1.2.1 Search for the best pruned tree

In most of the previous boosting algorithms (Chen and Guestrin 2016; Friedman 2001; Kokel et al. 2020), except *num_trees*, *max_depth* and *num_leaves* are two key hyperparameters which affect the classifier's performance significantly. Manually tuning the hyperparameter combinations is a heavy task and it is hard to find the best parameter combinations for different datasets. Therefore, we propose a novel function called resampling weighted pruning to automatically prune redundant leaves and produce robust tree models, where weights are used to establish a relationship between the pruning and boosting practices.

First, we define the original learning sample set as $L$, and randomly divide it into $V$ subsets, $L_v, v = 1, \ldots, V$, then, the training set of each subset is $L^{(v)} = L - L_v$. $T_{max}$, represents the tree comes from the original set $L$, and we build a complete tree in each subset $L_v$. The decision trees' cost function is defined:

$$
\begin{aligned}
Gini(\{T\},\{w\}) &= \sum_{|T|}\left[Gini(T)\right] \\
&= \sum_{|T|}\left[1 - \sum_{c=1}^{C} P_c^2\right] \\
&= \sum_{|T|}\left[1 - \sum_{c=1}^{C}\left(\frac{\sum_{i_o} w_{m,i_c}}{\sum_i w_{m,i}}\right)^2\right]
\end{aligned}
\quad (1)
$$

Where $|\tilde{T}|$ means leaves' number, $C$ represents the class number, the sample of class $c$ is defined as $i_c$. The loss of the trees is calculated by summarizing the gini impurity of all the leaves. Because each leaf node contains only the same class samples, the loss of an entire tree is generally zero. However, in the pruning process, $Gini(\{T\},\{w\})$ will increase when the samples of the pruning nodes are combined into their parent node. Since $Gini(\{T\},\{w\})$ always favors large trees, it is not the best method to select a pruned tree. Therefore, we add a penalty term, regularization parameter $\alpha$ and the tree leaves $|\tilde{T}|$ to the cost function. The new equation is shown as follows:

$$R_\alpha(T) = Gini(\{T\},\{w\}) + \alpha \cdot |T| \qquad (2)$$

When $\alpha$ is constant and $|\tilde{T}|$ decreases with pruning, the penalty term is the benefit of a smaller tree.

Here, $R_\alpha(T - T_t) - R_\alpha(T)$ defines the variation in the cost function, where $T$ means the complete tree, $T_t$ means the branch with the node at $t$, so the tree pruned at node $t$ should be $T - T_t$. Next, $R_\alpha(T - T_t)$ is equivalent to the branch at node $t$, so as to calculate the cost of the pruning on the internal nodes.

$$\begin{aligned} & R_\alpha(T - T_t) - R_\alpha(T) \leq 0 \\ \Rightarrow\ & R_\alpha(t) - R_\alpha(T_t) \leq 0 \\ \Rightarrow\ & Gini(\{t\},\{w\}) + \alpha - Gini(\{T_t\},\{w\}) - \alpha|\tilde{T_t}| \leq 0 \\ \Rightarrow\ & \frac{Gini(\{t\},\{w\}) - Gini(\{T_t\},\{w\})}{|T_t| - 1} \leq \alpha \end{aligned} \qquad (3)$$

Where,

$$g(t) = \frac{Gini(\{t\},\{w\}) - Gini(\{T_t\},\{w\})}{|T_t| - 1} \qquad (4)$$

When $\alpha \geq g(t)$, the cost value will decrease, and the branch $T_t$ will be pruned. The order in which we pruning the branches begins like this: first, set $\alpha = argmin\ g(t)$ to find the branch, and prune the branch, then repeat the process until the tree is left with the root node. This provides a sequence of pruned trees $\{T_\alpha^{(v)}, \alpha = 0, \ldots\}$ with the associated cost-complexity parameter $\alpha$.

For $\alpha$, we use the pruned tree $T_\alpha^{(v)}$ to estimate the $v - th$ subset and obtain the following training error:

$$TE_\alpha^{(v)} = \frac{\sum_{i_{miss}} w_{m,i_{miss}}^{(v)}}{\sum_i w_{m,i}^{(v)}} \qquad (5)$$

where $i_{miss}$ means the index of the misclassified sample's weight, $w_{m,i}^{(v)}$ means the sample weight of the test set $L_v$ and $TE_\alpha^{(v)}$ means the misclassified rate of set $L_v$. Therefore, the average misclassified rate of $v$ is obtained as follows:

$$TE_\alpha = \frac{1}{V}\sum_{v=1}^{V} TE_\alpha^{(v)} \qquad (6)$$

Meanwhile, we denotes $\alpha^* = argmin_\alpha TE_\alpha$, and through pruning $T_{max}$ till $R_{\alpha^*}(T_{max})$ reaches the minimum, we obtained the best pruned tree.

3.1.2.2 Inverse boosting structure

As shown in the 4th and 5th steps of Algorithm 1, Adaboost adopts the training error $\varepsilon_m$ as the boosting coefficient of the weak leaner to update the estimator's weights and the sample weights. However $\varepsilon_m$ is not a suitable criteria for the pruned trees. Actually, we should give the pruned trees higher estimated weights when they have lower training errors. Therefore, we suggest using a novel boosting structure, which associates the classification outcome with the pruned trees accordingly.

Firstly, we fit the training data $L$ into a complete decision tree and prune it to obtain the best tree structure. Since $TE_\alpha$ truly reflects the result of the pruned tree, we apply $TE_\alpha$ instead of $\varepsilon_m$ as the evaluation criteria. The weight of the estimator would be updated as follows,

$$\varphi_m = ln\frac{1-TE_\alpha}{TE_\alpha} \text{ s.t. } TE_\alpha < \frac{1}{2} \tag{7}$$

The training process stop when $TE_\alpha \geq \frac{1}{2}$, because the current estimator cannot maintain the classification performance at all times.

Next, we propose an inverse structure to update the sample weight. In each iteration, we treat the misclassified samples as 'intractable' items, which may affect the judgement and the robustness of the pruned tree. Therefore, inspired by Tong et al. (2019), the sample weight can be updated:

$$w_{m+1,i} = \frac{w_{mi}}{Z_m} exp\left(\varphi_m I\left(G_m(X_i), y_i\right)\right), i=1,2,\ldots N \tag{8}$$

$$D_{m+1} = \left(w_{m+1,1}, w_{m+1,2}\ldots\ldots w_{m+1,N}\right), m=1,2,\ldots M \tag{9}$$

In Eqn. (8), $Z_m$ is a normalization factor

$$Z_m = \sum_i w_{mi} exp\left(\varphi_m I\left(G_m(X_i), y_i\right)\right) \tag{10}$$

$G_m(X_i)$ is the estimated value of $X_i$ by the pruned tree $G_m$, $y_i$ means the ground-truth of sample $X_i$, and $I(G_m(X_i), y_i)$ is defined:

$$I\left(G_m\left(X_i\right), y_i\right) = \begin{cases} 1, G_m\left(X_i\right) = y_i \\ -1, G_m\left(X_i\right) \neq y_i \end{cases} \quad (11)$$

The weight of the misclassified samples is decreasing, while the weight of the classified samples is increasing, which reduces the influence of the "intractable" items, which help to optimize the tree building in the subsequent iterations. Then, the next estimator is trained in the dataset $L$ with a new weight distribution $D_{m+1}$. The training process will execute iteratively until meet the hyperparameter *num_tree* M. The final integrated classifier is:

$$G_{final}(X) = sign\left(\sum_{m=1}^{M} \varphi_m G_m(X)\right) \quad (12)$$

The whole procedure is shown in Algorithm 2.

3.1.3. Hyperparameter Optimization

Hyperparameters determine the pre-defined characteristics of a classifier and hyperparameter should be set before the training process begins. Our novel classifier IBPT has few hyperparameters and here we use the Grid search method to tune the hyperparameters of IBPT: number of pruning trees, sample minimum number per leaf and iteration number of the pruning process. More specifically, the minimum number of the samples per leaf controls the pre-defined complexity of the decision trees and the final depth of the trees will be determined by our pruning methods. We create different combinations against these three parameters and use the Grid search method to search for the optimal hyperparameter list of IBPT. The performance of the trained classifiers is compared using five-fold cross-validation: the training data is divided randomly into five subsets and each time we will use four subsets to train a new IBPT classifier with the remaining data as the validation set. Then, comparing the validation results of different IBPT models (using the Grid Search method), the best model is identified and its hyperparameters are selected for the final model in the testing stage.

---

**Algorithm 2** Inverse boosting pruning trees based algorithm.

**Require:** *M*-Trees' number, *N*-Samples number, *L*-Learning samples, and *V*-Folds.

1: **function** BEST PRUNED SUBTREE (*L*, *V*, $D_m$).
2:  Split the learning samples *L* into *V* folds, $L_v, v = 1, 2, \ldots, V$, and grow a max tree $T_{max}$ on *L*.
3:  Test sample set $L^v = L - L_v$.
4:  **for** $v \in [1, V]$ **do**
5:   Fit a decision tree to $L_v$ training samples.
6:   Subtree sequence $\{T_\alpha^{(v)}, \alpha = 0, \ldots\} \leftarrow R_\alpha(T - T_t) - R_\alpha(T) \leq 0$

   Recursively repeat till the tree only has root nodes: 1. Calculate $g(t)$ using Eq. (4). 2. Set $\alpha = argmin\, g(t)$ and prune the branch $T_t$.
7:   Calculate $TE_\alpha^{(v)}$ by Eq. (5).
8:  **end for**
9:  Compute $TE_\alpha \leftarrow \frac{1}{v} \sum_{v=1}^{V} TE_\alpha^{(v)}$
10: Define $\alpha^* \leftarrow argmin_\alpha TE_\alpha$.
11: The best pruned tree $G_m(X)$ is obtained by pruning $T_{max}$ till $R_{\alpha^*}(T_{max})$ becomes minimal.
12: **return** $G_m(X), TE_\alpha$.
13: **end function**
14:
15: **function** INVERSE BOOSTING(*L*, *V*, *M*, *N*)
16: Initialise sample weight distribution $D_m = (w_{mi} \ldots), m = 1, 2, \ldots M, i = 1, 2, \ldots N$ and set each sample weight $w_{mi}$ to $\frac{1}{N}$.
17: **for** $m \in (1, M)$ **do**
18:  $G_m(X), TE_\alpha \leftarrow$ Best Pruned Subtree (*L*, *V*, $D_m$).
19:  Update the estimator weight using Eq. (7).
20:  Update each sample's weight $w_{m,i}$ using Eqs. (8) and (9).
21:  Preserve Di for the next iteration
22: **end for**
23: **return** Final ensemble classifier $G_{final}(X) \leftarrow sign\left(\sum_{m=1}^{M} W_t^m G_m(X)\right)$.
24: **end function**

3.1.5. Feature Importance

The feature importance: The importance of a feature is computed as the (normalized) reduction of the errors brought by that feature. It is also known as the Gini importance. The single node importance *NI* is defined as:

$$NI = Gini\left(\{Node_{split}\},\{w\}\right) - Gini\left(\{Node_{left}\},\{w_{left}\}\right) \\ - Gini\left(\{Node_{right}\},\{w_{right}\}\right) \tag{13}$$

Where $Node_{split}$ is the splited non-leaf node in the decision tree, $Node_{right}$ and $Node_{left}$ are the right and left children nodes of $Node_{split}$.

The importance for each feature on a decision tree is then calculated as:

$$feature\,importance = \frac{1}{M}\sum_{m}^{M}\frac{\sum_{f}^{F}NI_{f}^{(m)}}{\sum_{k}^{K}NI_{k}^{(m)}} \tag{14}$$

So M is the number of the trees in IBPT, F is the number of non-leaf nodes which employ the target feature to split data and *K* is the total number non-leaf in the *m-th* tree.

*3.2. Performance Evaluation*

In this study, we formulate the task of earthquake forecasting as a binary class classification problem and use eight performance measures, namely, Matthews correlation coefficient (MCC), Hanssen–Kuipers discriminant (R score), the Area Under the Curve (AUC), Specificity, Sensitivity, Accuracy and Precision and the Area Under the Recall-Precision Curve (AURPC), and to test the effectiveness of these measures.

Area Under the Curve is the area under the receiver-operating characteristic (ROC), it is a plot of true positive rate (TPR) against false positive rate (FPR). In practice, AURPC is also often used to test the effectiveness, so AURPC can be a good option for the area under the curve (Davis and Goadrich 2006).

The Accuracy (ACC) is defined as:

$$ACC = \frac{TP+TN}{TP+TN+FP+FN} \tag{15}$$

The Sensitivity (TPR) is defined as:

$$TPR = \frac{TP}{TP+FN} \tag{16}$$

The Specificity (TNR) is defined as:

$$TNR = \frac{TN}{TN+FP} \tag{17}$$

The Precision is defined as:

$$PR = \frac{TP}{TP+FP} \tag{18}$$

The Hanssen–Kuipers discriminant (R score) (Hanssen and Kuipers 1965) is defined as:

$$R\ score = \frac{TP \times TN - FP \times FN}{(TP+FN)(FP+TN)} \tag{19}$$

Except for the metrics mentioned above, which stress on positives, it also used Matthews Correlation Coefficient (MCC) (Matthews 1975). This coefficient is a balanced measure, and it can measure the correlation between the expected class and the obtained class. MCC is calculated as:

$$MCC = \frac{TP \times TN - FP \times FN}{\sqrt{(TP+FP)(TP+FN)(TN+FP)(TN+FN)}} \tag{20}$$

where *TP* means the true positives, *TN* means the true negatives, *FP* means the false positives and *FN* means the false negatives, respectively.

**4. Results and Discussion**

*4.1 Comparison of results with different machine learning methods*

As shown in Table S5 and Table S6, all the benchmarking methods were used to construct models for forecasting, based on the two datasets with the generated features. No significant unbalance was found in the training and testing datasets, suggesting the credibility and stability of the forecasting models. The performance metrics of MCC, R score and AUC were used to evaluate the performance of all the methods on the testing datasets. For Dataset I, the MCC of each method ranges from 0.4903 to 0.6581, the R score of each method ranges from 0.4643 to 0.6429, and the AUC of each method ranges from 0.5829 to 0.8718. We found that IBFT was the top performer for Dataset I (MCC =0.6581, R score = 0.6429 and AUC = 0.8718). The ROC curves of the methods on Dataset I are shown in Figure 2.

For dataset II, IBPT was still the top performer (MCC =0.5958, R score = 0.5942 and AUC = 0.8683). The ROC curves of the methods on dataset II are shown in Figure 3. IBPT appears to be robust in accuracy on the two datasets. The MCC and accuracy of IBPT were the best on both the datasets. RF performed quite differently over these two datasets, though.

Furthermore, we observed that IBPT performed better for the earthquakes with larger magnitudes (i.e., Dataset I). This can be explained by the fact that the features selected for larger earthquakes are more supportive in discriminating pre-earthquake perturbations. RF achieves the second-best accuracy in Dataset I (accuracy = 0.7679) and Dataset II (accuracy = 0.785). This suggests that a tree-based classifier is capable of producing better performance. Although MLP and CNN are constructed with four layers, e.g. fully connected and convolutional layers, their performance is worse than those of RF and GBM. This might be due to the fact that deep learning architectures require significantly large training sets (a large number of earthquakes) for system optimization and this is not available in the current research domain with very limited resources.

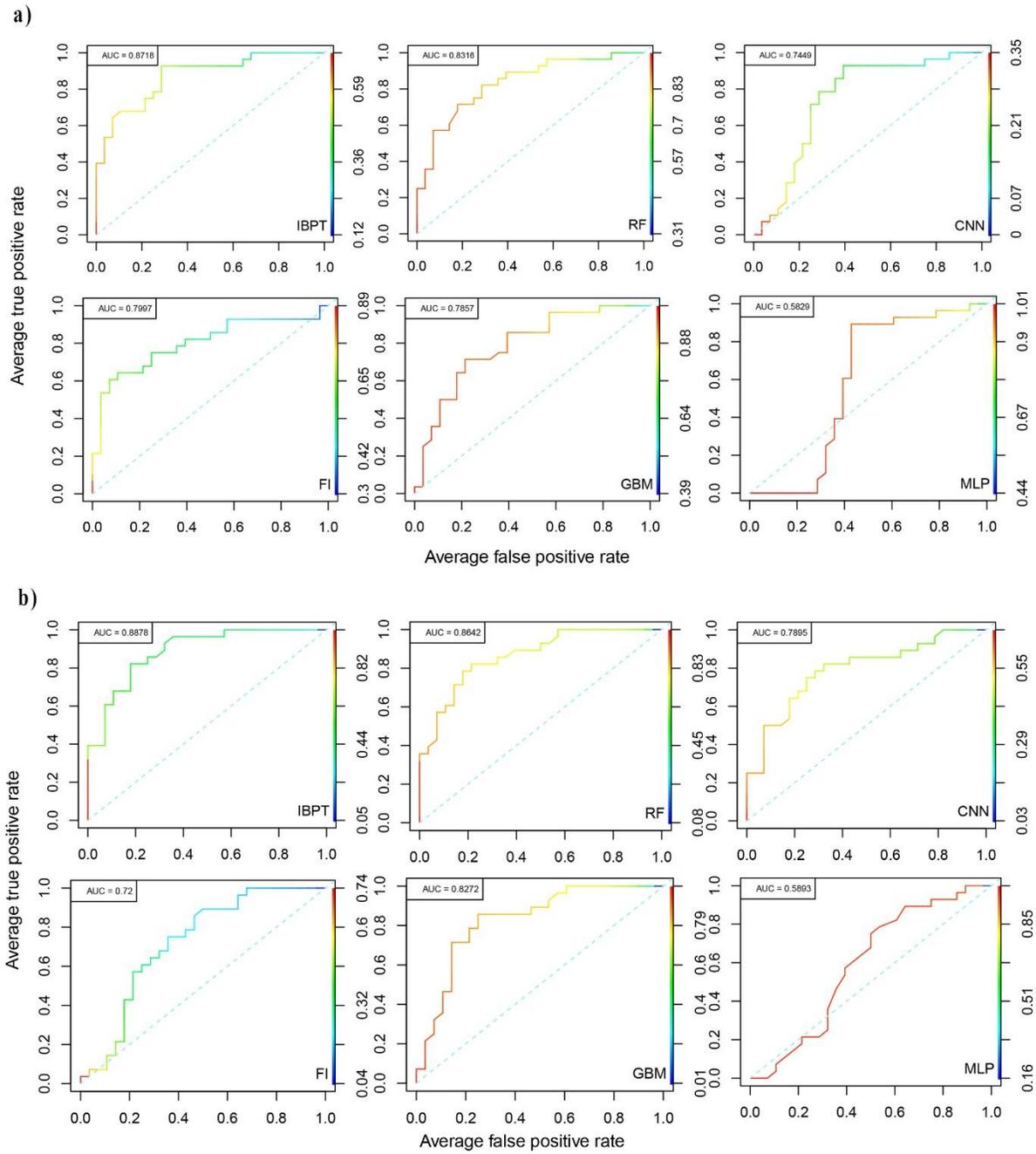

Figure 2 The ROC curves of the six best-benchmarking methods on the satellite dataset of earthquakes of magnitude 7 or greater with the proposed features a) with aftershocks (Dataset I) b) with aftershocks dropped (Dataset V).

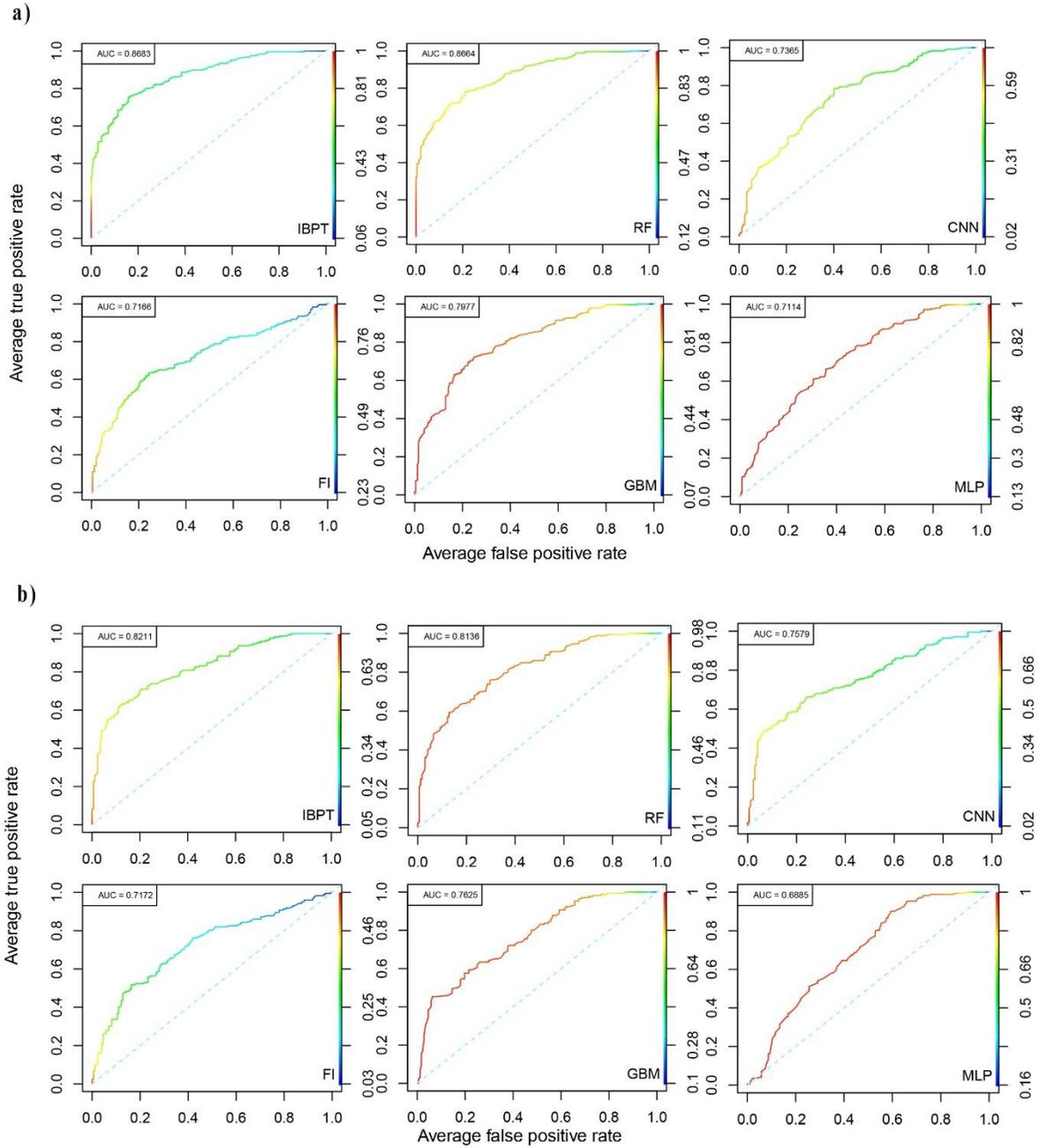

Figure 3 The ROC curves of the six best-benchmarking methods on the satellite dataset of earthquakes of magnitudes between 6 and 7 with the proposed features a) with aftershocks (Dataset II) b) with aftershocks dropped (Dataset VI).

*4.2 Comparison between different features*

As illustrated in Table S7 and Table S8, DataSet III, and DataSet IV were used to construct the models for forecasting, based on the two datasets (compared with the proposed "sliding window" features based on DataSet I and II). In general, we discover that the datasets with the proposed features (DataSet I and II) lead to better classification performance than the datasets with the standard features (DataSet III and IV) for all the classifiers used. The ROC curves of the methods for DataSet III and IV are shown in Figure 4 and Figure 5 for performance comparison, respectively.

In further analysis, we use IBPT as an example to analyse the experimental results, and generate four datasets based on the spatial and temporal features.

As seen in Table S9, all the benchmarking datasets were used to compare the results of using different features by IBPT. We observe that the datasets with the proposed "sliding window" features lead to better classification outcomes than the datasets with the standard features: The MCC of IBPT on the two datasets with the proposed features are 0.6581 and 0.5958, respectively, and the MCC on the two datasets with the standard features are 0.6429 and 0.5258, respectively. From this observation, we interpret that the datasets with the proposed features enable the earthquake forecast to achieve better accuracy than those with the standard features. This is because the way of generating the proposed features by a "sliding window" style that covers 5 days (or so) observation data extracts sufficient information while reducing data redundancy. As shown in Figure 6, total integrated column ozone burden, outgoing longwave radiation flux (NOAA) and retrieved total column CO are the most important features rendered by the trained IBPT model when it is used to discriminate the earthquake and non-earthquake data.

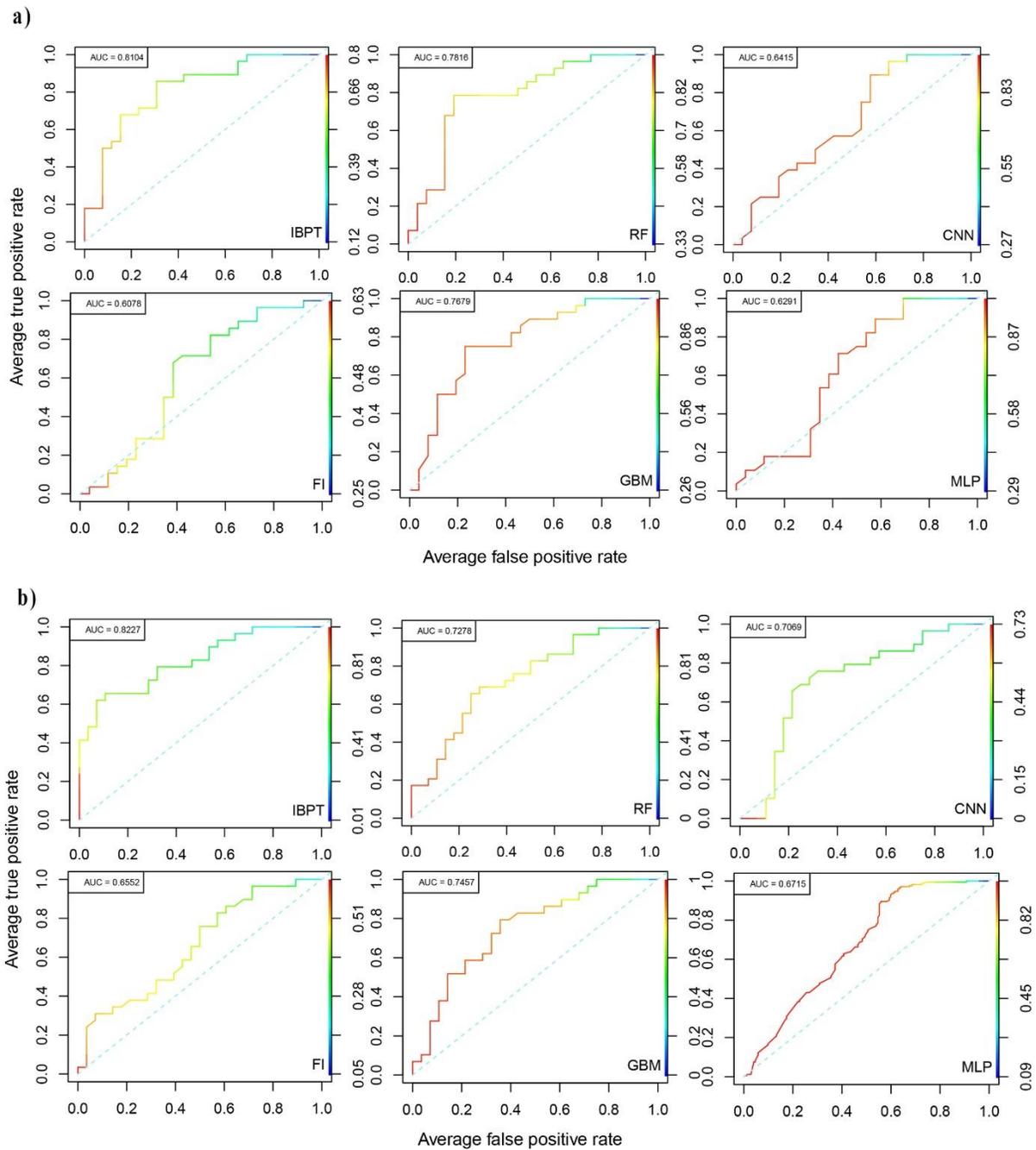

Figure 4 The ROC curves of the six best-benchmarking methods on the satellite dataset of earthquakes of magnitude 7 or larger with the standard features a) with aftershocks (Dataset III) b) with aftershocks dropped (Dataset VII).

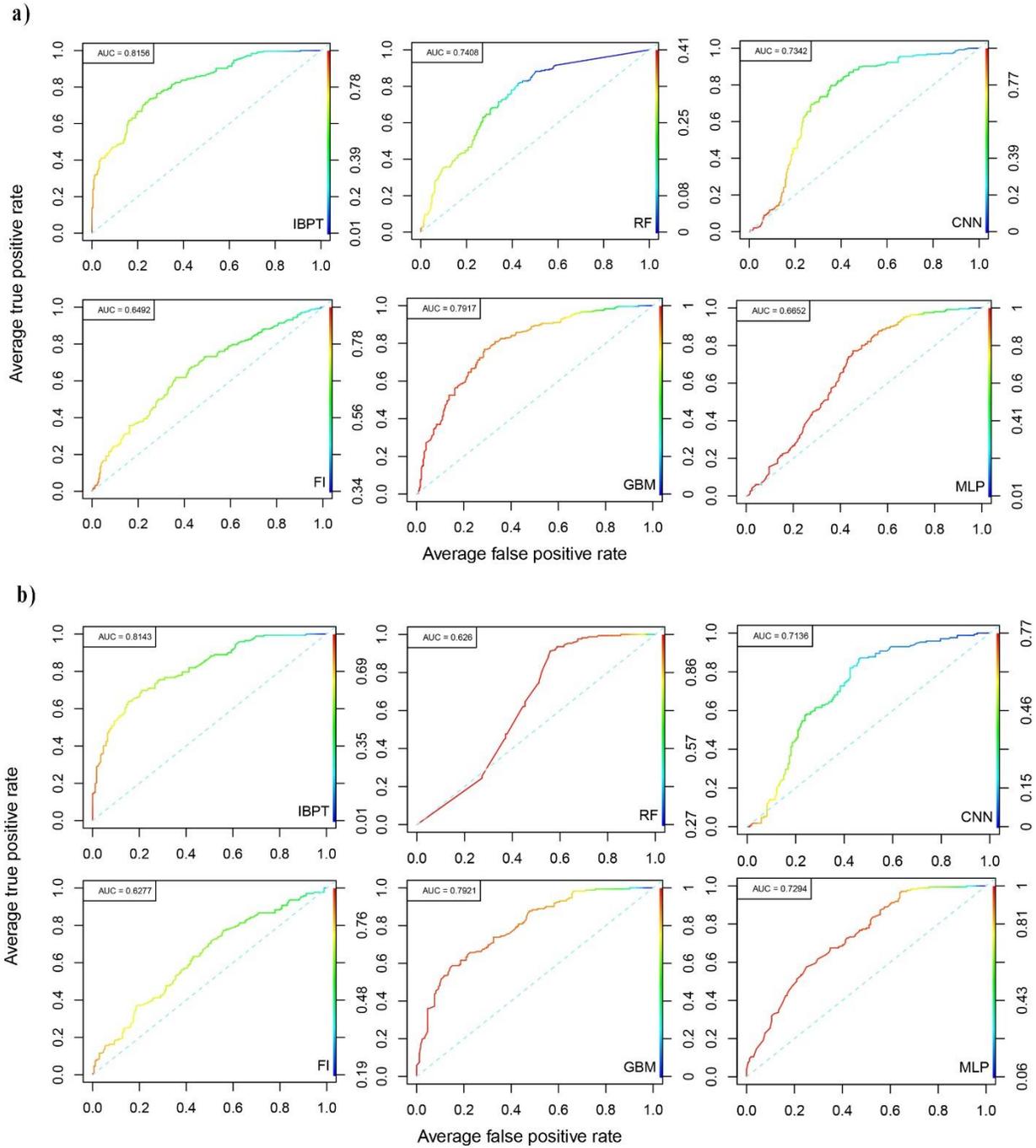

Figure 5 The ROC curves of the six best-benchmarking methods on the satellite dataset of earthquakes of magnitudes between 6 and 7 with the standard features a) with aftershocks (Dataset IV) b) with aftershocks dropped (Dataset VIII).

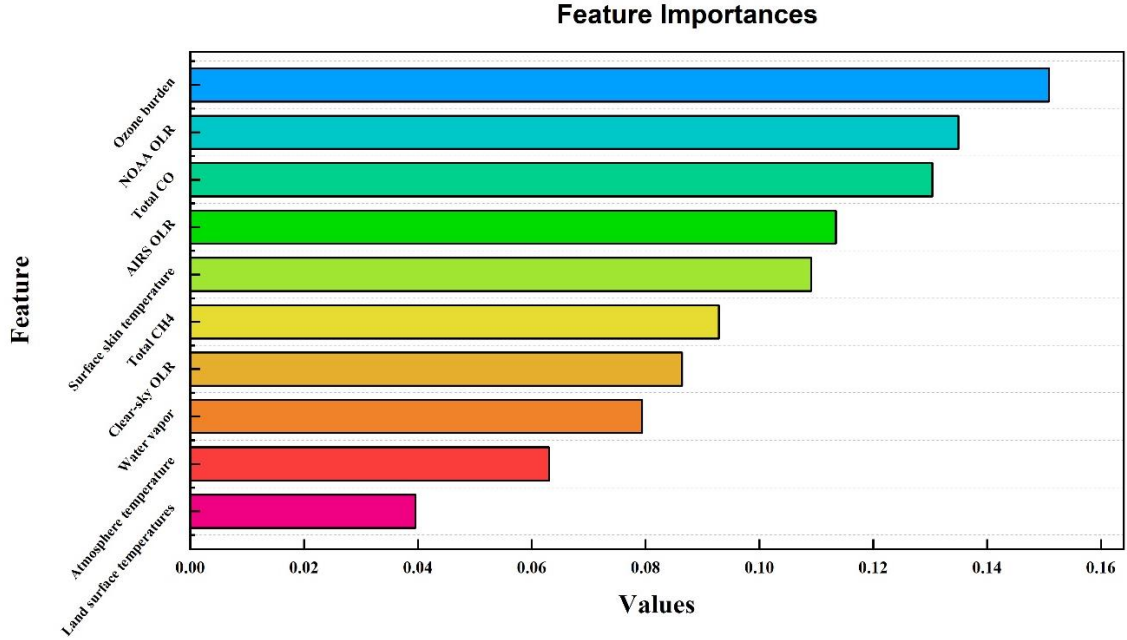

Figure 6 Features' importance is evaluated by the proposed IBPT model on the satellite dataset of earthquakes of magnitudes with the proposed features. The colours shown here are used merely for better display quality.

Table S10 and Table S11 present the forecasting performance of the six best-benchmarking methods on the datasets of the proposed features with non-overlapping windows. From Table S10, the MCC ranges from 0.3299 to 0.6075 and the accuracy ranges from 0.6429 to 0.8036 for different methods. It remains true that IBPT was the top performer for the dataset of earthquakes of magnitude 7 or greater with non-overlapping windows. It has also been found that IBPT outperforms all the selected baselines for the dataset of earthquakes of magnitude between 6 and 7 with non-overlapping windows in Table S11. The ROC curves of the methods are shown in Figure 7 and Figure 8 for performance comparison, respectively. IBPT was still the top performer for the two datasets. Besides, in most cases, the models work better on datasets when overlapping windows are used for generating time series (i.e., DataSet I and DataSet II) compared to the cases with non-overlapping windows used (i.e., DataSet I-nonoverlap and DataSet II-nonoverlap). A possible reason is that while the features based on non-overlapping windows are less correlated (which have positive effects on the performance), the size of training dataset (generated based on the same raw dataset) are smaller (which have the negative effects on the performance). Besides, IBPT was still the top performer for the two newly considered datasets.

*4.3 Considering the aftershock effect*

The aftershocks may play an active role on earthquake forecasting. To demonstrate the aftershock effect, we have ruled out the aftershocks and carried out a comparative study. So, it is necessary to delete the data corresponding to aftershocks from the list of earthquakes (Yan et al. 2017). In our work, we associated an area of 2° × 2° centered on the epicenter for all earthquakes in the list, and to get the result. Practically, we processed the list in the following chronological order: First, select an earthquake (given earthquake) in the list, the setting feature is the time of the earthquake and its related region. Then, from the corresponding data list of the system, to remove any other earthquake occurred in the related area within 30 days after the given earthquake occurred. In our research framework, it is considered that 30 days is the maximum period of anomaly before the earthquake (we set the temporal window to be 30 by default in our study). Finally, we dropped 390 aftershocks. In addition, the data corresponding to the days of the aftershocks are deleted. After these operations, 981 independent earthquakes still remain in the list.

According to Table S5, the proposed method IBPT is the top performer for the satellite dataset of earthquakes (with aftershocks dropped) of magnitude 7 or greater with the proposed features (Dataset V) (MCC =0.6429, R score = 0.6429 and AUC = 0.8878). The ROC curves of the methods for Dataset V are shown in Figure 2b. For Dataset VI (the satellite dataset of earthquakes (with aftershocks dropped) of magnitudes between 6 and 7 with the proposed features), IBPT is still the top performer (MCC = 0.5258, R score = 0.5058 and AUC = 0.8211). The ROC curves of the methods for Dataset VI are shown in Figure 3b. The MCC and accuracy of IBPT were the best on both the datasets.

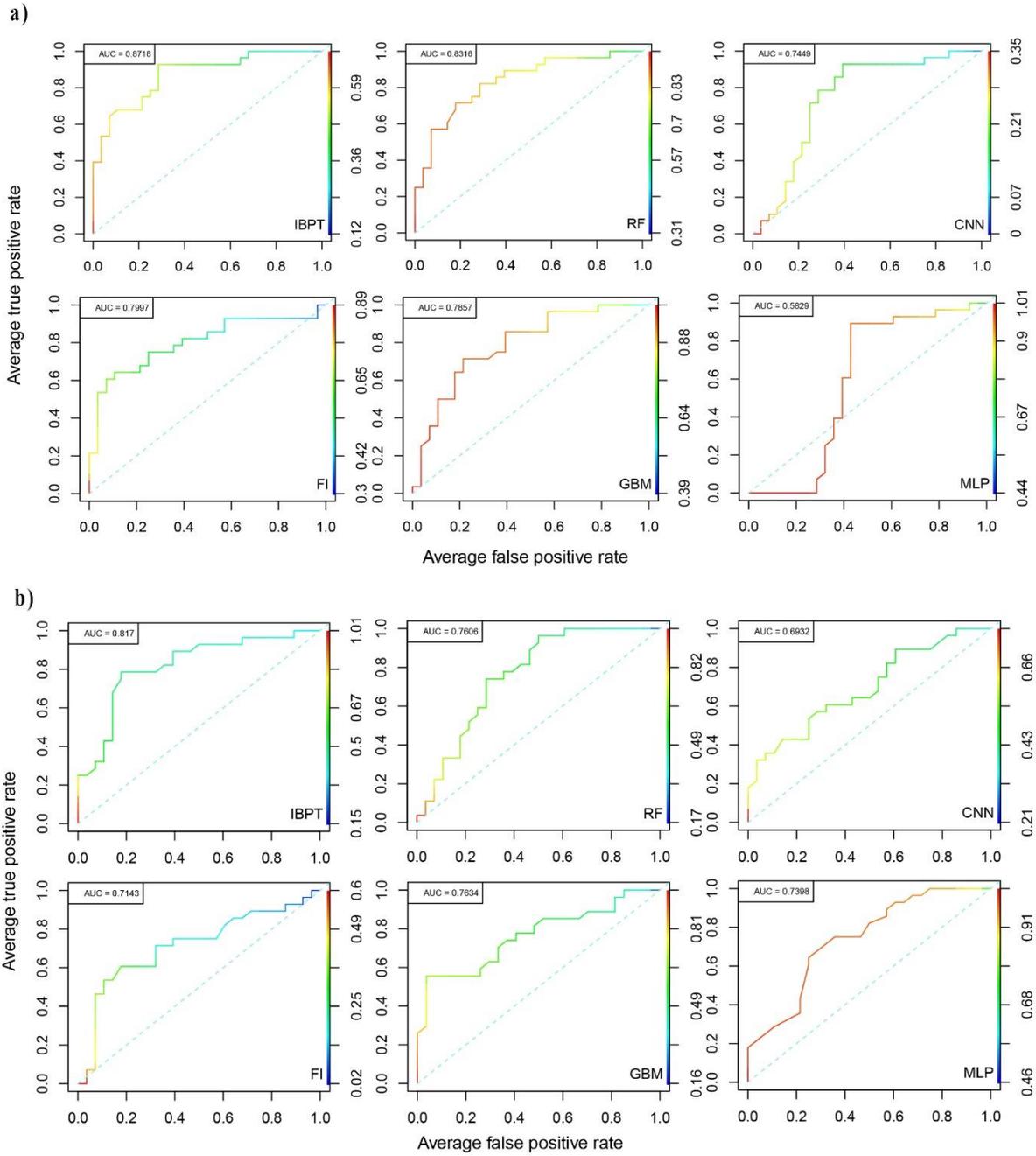

Figure 7 The ROC curves of the six best-benchmarking methods on the satellite dataset of earthquakes of magnitude 7 or greater with the proposed features a) with overlapped data (Dataset I) and b) with non-overlapped data (Dataset I-nonoverlap).

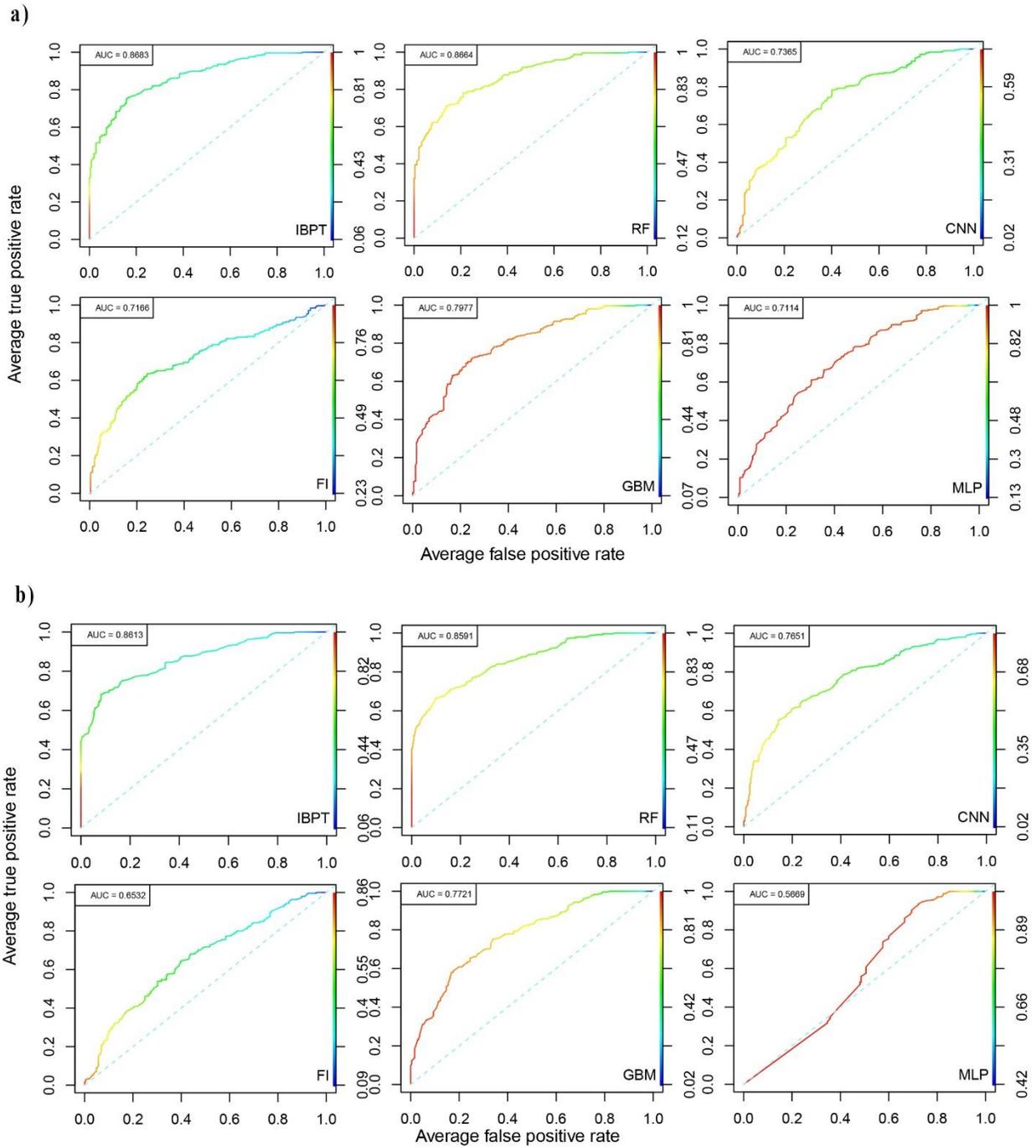

Figure 8 The ROC curves of the six best-benchmarking methods on the satellite dataset of earthquakes of magnitudes between 6 and 7 with the proposed features a) with overlapped data (Dataset II) b) with non-overlapped data (Dataset II-nonoverlap).

As illustrated in Figure 4b and Figure 5b, DataSet VII (the satellite dataset of earthquakes (with aftershocks dropped) of magnitude 7 or greater with the standard features) and DataSet VIII (the satellite dataset of earthquakes (with aftershocks dropped) of magnitudes between 6 and 7 with the standard features) were used to construct the models for forecasting, based on the standard features (compared with the proposed "sliding window" features based on DataSet V and DataSet VI). In general, we discover that the datasets with the proposed features (Dataset V and VI) lead to better classification performance than the datasets with the standard features (DataSet VII and VIII) for all the classifiers used. The ROC curves of the methods for DataSet VII and VIII are shown in Figure 4b and Figure 5b for performance comparison, respectively. IBPT was still the top performer for the two datasets.

In general, although we dropped 390 aftershocks, our work shows that the proposed IBPT framework outperforms the chosen state of the art methods, and becomes the top performer for all the benchmarking datasets. Our work also further proves that the proposed IBPT model in combination with the proposed features performs better than the methods with the standard features, aftershocks had no effect on our result.

*4.4 Considering different temporal windows*

We have observed that satellite data with a temporal window of 30 days (DataSet V) has good forecasting precision. In order to investigate whether or not our proposed method is capable to predict earthquake with different temporal windows, satellite datasets with temporal windows of 05 days (Dataset IX), 10 days (Dataset X), 15 days (Dataset XI), 20 days (Dataset XII) and 25 days (Dataset XIII) have been generated (shown in Table S12).

Figure S5a provides the ROC curve of the six datasets with different temporal windows. Table S13 presents the forecasting performance with different temporal windows using IBPT. From Table S13, the MCC ranges from 0.3953 to 0.6429 and the accuracy ranges from 0.6727 to 0.8214 on different datasets. It remains true that by reducing the days of the temporal window, the performances decrease by about 0.24 for MCC and 0.14 for accuracy. That is, the proposed model's performance is worse than that if we reduce the days of the temporal window. Based on these results, we conclude that the choice of the temporal window size is influencing, to a certain extent, the performance of the proposed model, by reducing its capability in predicting earthquakes. Although our proposed method is capable to predict earthquake with different temporal window sizes, it gives the best performance on the dataset with our initial selection of the temporal window of 30 days.

*4.5. Considering Different spatial windows*

Although satellite data with a spatial window with its center at the epicenter and a deviation of 3° (DataSet V) shows a strong capability in earthquake forecasting, satellite datasets of the spatial window with its center at the epicenter and a deviation of 1° (DataSet XVIII), 2° (DataSet XIX), 4° (DataSet XX) and 5° (DataSet XXI) have been generated (Table S12) in order to further find the optimal spatial window.

Table S14 presents the prediction performance with the five datasets of different spatial windows using IBPT. From Table S14, the AUC on each dataset ranges from 0.7389 to 0.8878 and the MCC ranges from 0.4388 to 0.6429. We discover that the best performance is with the dataset with its center at the epicenter and a deviation of 3° (DataSet V, with AUC of 0.8878 and MCC of 0.6429), and by using different distances of the spatial window, the performance of AUC and MCC decreases by about 16.7% and 31.7%, respectively. From this observation, we conclude that although the IBPT model is capable of forecasting earthquake with different spatial window sizes, the dataset with its center at the epicenter and a deviation of 3° enable earthquake forecasting using satellite data to achieve better performance than those with other distance of the spatial window. Figure S5c provides the ROC curves of the five datasets (including DataSet V) with different spatial windows.

*4.6. Considering unbalanced dataset*

As the actual earthquake problem is always highly unbalanced, where non-earthquakes instances are always higher as compared to earthquakes. In order to provide the realistic performance overview, we try our proposed method on unbalanced dataset in this section. To investigate whether or not our proposed method is capable to predict earthquake with unbalanced datasets, satellite dataset with the positive to negative ratio of 1:2 (Dataset XIV), 1:5 (Dataset XV), 1:10 (Dataset XVI) and 1:15 (Dataset XVII) have been generated (shown in Table S12).

Table S15 illustrates the proposed method's performance on the five datasets. As is shown in, the method has similar performance over the six datasets, e.g., the MCC of the proposed method on all the five datasets is around 0.62 (ranging from 0.6145 to 0.6429), the accuracy of the proposed method on all the five datasets is around 0.83 (ranging from 0.8214 to 0.8588). Figure S5b shows the ROC curves, we also observe a similar tendency that the performance of our method on the five datasets, suggesting that our method provides satisfactory performance for earthquake forecasting on the unbalanced datasets. Although the five unbalanced datasets are quite different, these results indicate that our method is not sensitive to the positive to negative ratio, and that our method can be used to predict earthquakes using unbalanced datasets, and provide good performance.

*4.7. Discussion*

We summarize the previous studies using machine learning for earthquake prediction and pre-earthquake perturbation analysis from the satellite data, shown in Table S16. Through the performance comparison among these studies, the result shows that among those methods, the IBPT method outperforms the others, i.e., it gives the best performance on all the benchmarking datasets. Moreover, since earthquake is a small probability event, the actual earthquake problem is always highly unbalanced. In this paper, we mainly use Matthew's Correlation Measure (MCC) to evaluate the performance. By comparison, the best MCC of our method IBPT can achieve 0.6581.

There are multiple sources of uncertainty associated with the infrared and hyperspectral satellite data and methods for earthquake forecasting. Uncertainty of data precision is caused by random effects in the data, and uncertainty of data accuracy is caused by systematic effects (Smith et al. 2015). Specifically, uncertainties of infrared OLR data can attribute to several factors, including the magnitudes and the degrees of persistency of the regional OLR diurnal and interannual variations (Gruber et al. 2007), the goodness of fit of the models (Moy et al. 2010), surface emissivity (Clerbaux et al. 2020), the AVHRR OLR's precision is with particularly large uncertainties in the deserts and elevated regions (Gruber et al. 2007). The uncertainty of surface skin temperature obtained from ARIS could be due to the short periods of the satellite based temperature records (Kang et al. 2015). Land surface temperatures uncertainties are affected by the methodologies for the surface retrieval and emissivity first guess (Hulley and Hook 2012). Pagano et al. (2020) give detailed discussion of measurement uncertainties of AIRS L1B radiances, and note that large uncertainty in the modules at low scene temperatures due to the larger polarization uncertainty, and the larger errors associated with the emissivity degradation in the shorter wavelength modules. Moreover, the wide variety of cloud complexity is an important factor in uncertainty for AIRS error estimation (Kahn et al. 2015; Wong et al. 2015). Furthermore, as parameter tuning of the proposed IBPT model is time-consuming and challenging, we cannot guarantee that optimized parameters were obtained for the models trained in each dataset, though most cases were covered through the grid search method employed in our study. Still, this introduced additional uncertainty to the earthquake forecasting of the proposed model.

One limitation of IBPT is that it has significant computational complexity because of the pruning methods. More specifically, in each iteration, the base tree of the IBPT needs to grow fully with the training data then iteratively prune leaf nodes from bottom to top. This process improves the fitting and generalization ability of the base trees but reduce its training speed. This problem can be handled by deploying more computing resources such as multiple CPUs or GPUs.

**5. Conclusions**

Hyperparametric optimization and cross-validation are used in our proposed system for earthquake forecasting, which allows us to find the best parameters for our model. In this way, we can perform model selection with high confidence, assuring that a robust model is selected and used. By comparison, our method IBPT improves by 16% in MCC (from 0.5657 to 0.6581) and more than 10% in R score (from 0.5357 to 0.6429) over the next-best CNN. It can be concluded that the proposed IBPT framework outperforms the chosen state of the art methods, and becomes the top performer for all the benchmarking datasets.Moreover, we could observe that infrared and hyperspectral satellite measurements in the circular region with its center at the epicenter and a radius of 3° and 30 days before the times of the shocks are more reasonable in earthquake forecasting. Our work also further proves that aftershocks had no effect on the result performed by the proposed IBPT model.

Our work also indicates that the proposed IBPT model in combination with the proposed features performs better than the methods with the standard features. The proposed time series based are able to help improve the accuracy of the earthquake forecasting task. It can significantly improve performance on the satellite dataset of earthquakes of magnitudes between 6 and 7, the MCC of IBPT with the proposed features improves by 13.3%, which shows the proposed IBPT scheme is effective to some extent on the datasets of a relatively large sample size. It can also be inferred from the feature importance analysis that total integrated column ozone burden, outgoing longwave radiation flux (NOAA) and retrieved total column CO are the most important features rendered by the trained IBPT model when it is used to discriminate seismic and non-seismic data.

Our work shows that the use of big satellite data analytics with machine learning is capable of successfully improving the likelihood of earthquake forecasting. In particular, earthquakes may be forecasted to some extent using the proposed IBPT framework with the proposed spatial and temporal features.

**Acknowledgements**

This work is supported in part by the National Key R&D Program of China under Grant No. 2018YFC1503505, and in part by the Special Fund of the Institute of Earthquake Forecasting, China Earthquake Administration under Grant 2020IEF0510 and Grant 2020IEF0705.

**Declaration of Interest**

The authors declare that they have no known competing financial interests or personal relationships that could have appeared to influence the work reported in this paper.

*Supplementary material for*

# Towards advancing the earthquake forecasting by machine learning of satellite data

**Pan Xiong** [1, 8], **Lei Tong** [3], **Kun Zhang** [9], **Xuhui Shen** [2, *], **Roberto Battiston** [5, 6], **Dimitar Ouzounov** [7],

**Roberto Iuppa** [5, 6], **Danny Crookes** [8], **Cheng Long** [4] and **Huiyu Zhou** [3]

[1] Institute of Earthquake Forecasting, China Earthquake Administration, Beijing, China

[2] National Institute of Natural Hazards, Ministry of Emergency Management of China, Beijing, China

[3] School of Informatics, University of Leicester, Leicester, United Kingdom

[4] School of Computer Science and Engineering, Nanyang Technological University, Singapore

[5] Department of Physics, University of Trento, Trento, Italy

[6] National Institute for Nuclear Physics, the Trento Institute for Fundamental Physics and Applications, Trento, Italy

[7] Center of Excellence in Earth Systems Modeling & Observations, Chapman University, Orange, California, USA

[8] School of Electronics, Electrical Engineering and Computer Science, Queen's University Belfast, Belfast, United Kingdom

[9] School of Electrical Engineering, Nantong University, Nantong, China

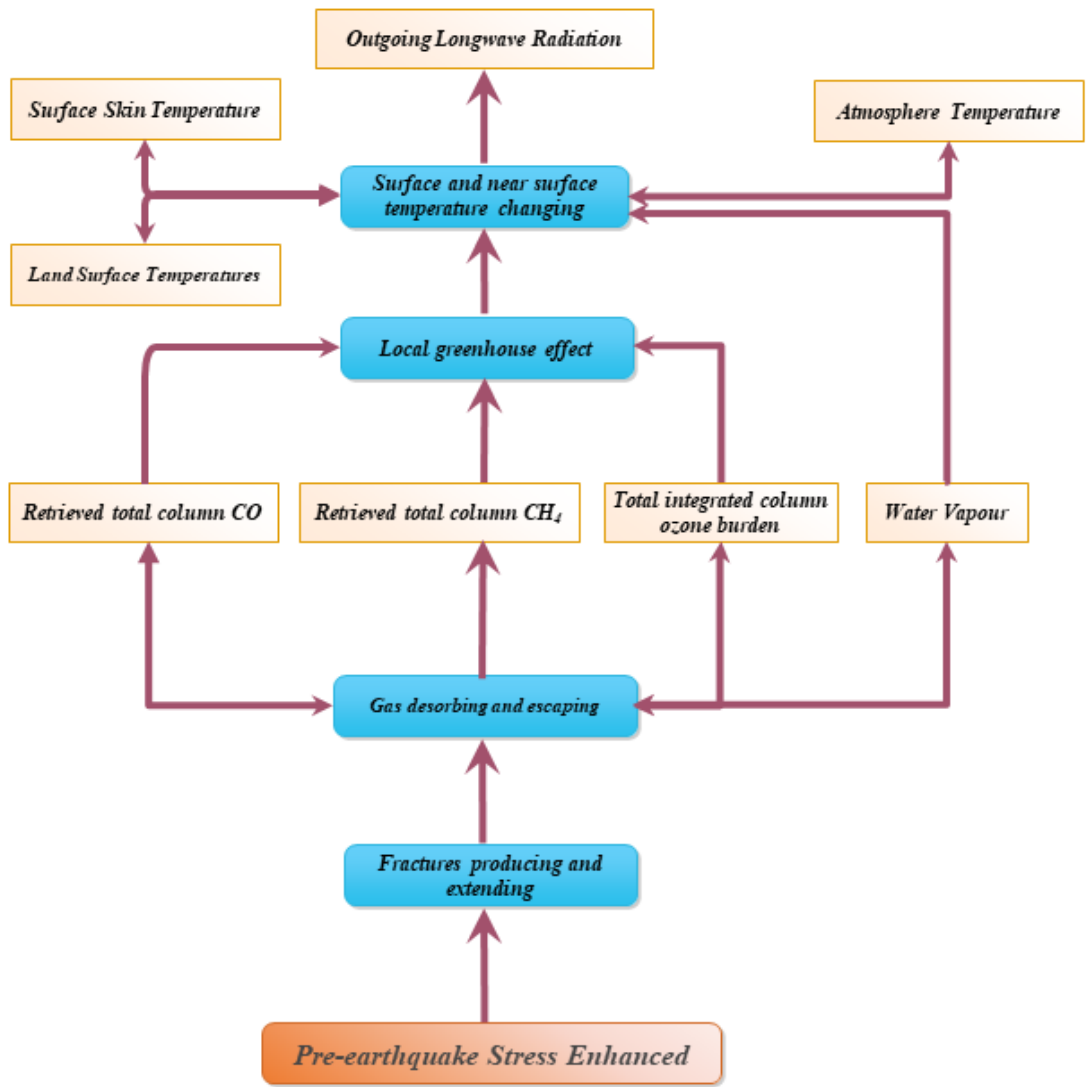

Figure S1 Simplified inherent relations between the selected ten parameters.

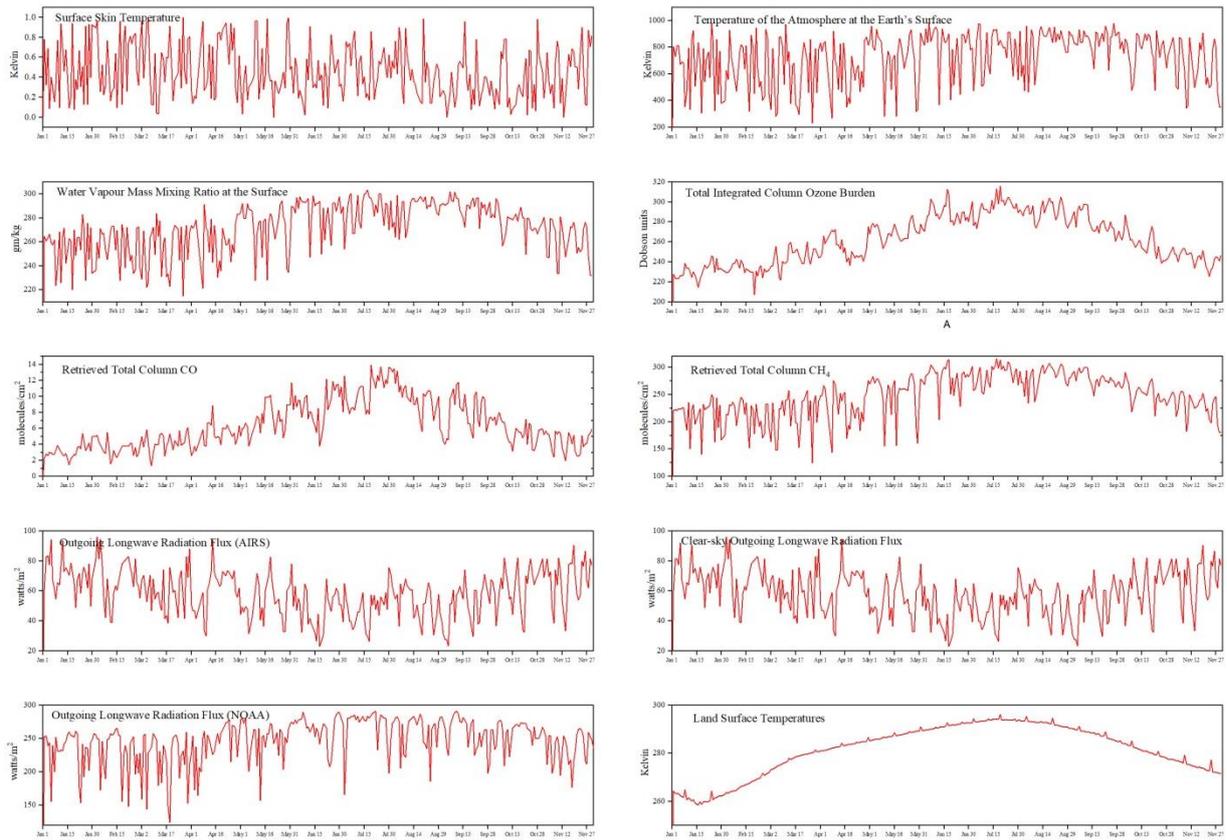

Figure S2 Example of ten standard features at latitude of 40.5 and longitude of 55.5 (2013).

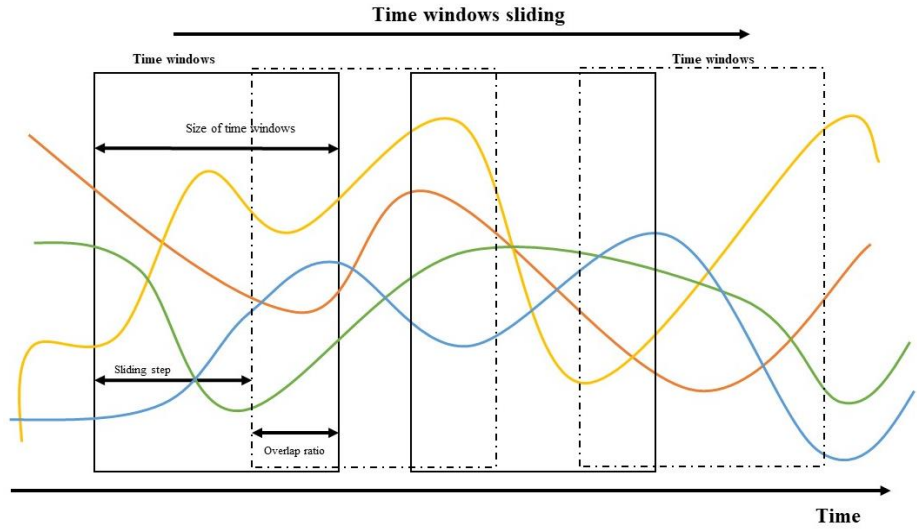

Figure S3 Sliding time window.

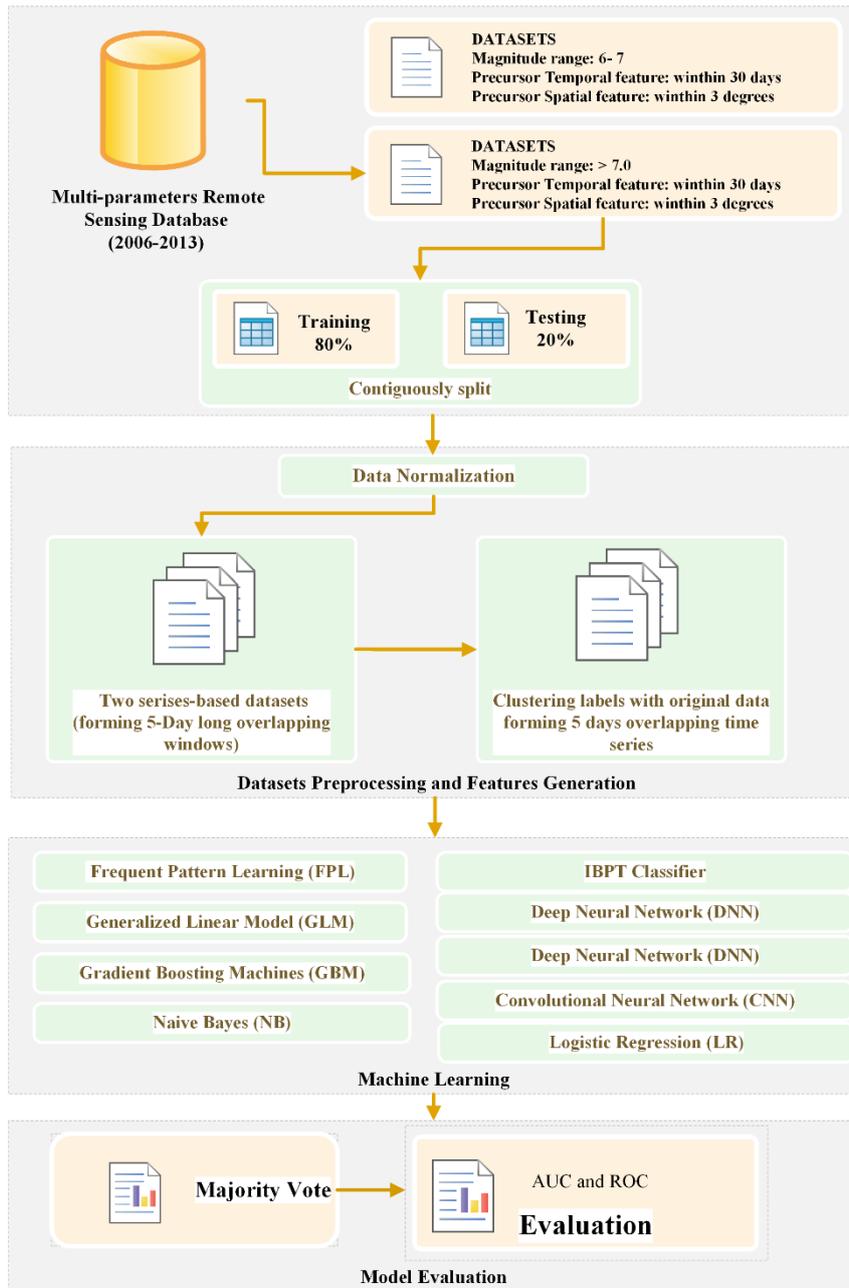

Figure S4 Schematic diagram of the methodology used to retrieve, divide, train and test the set of machine learning methods used to perform the proposed study

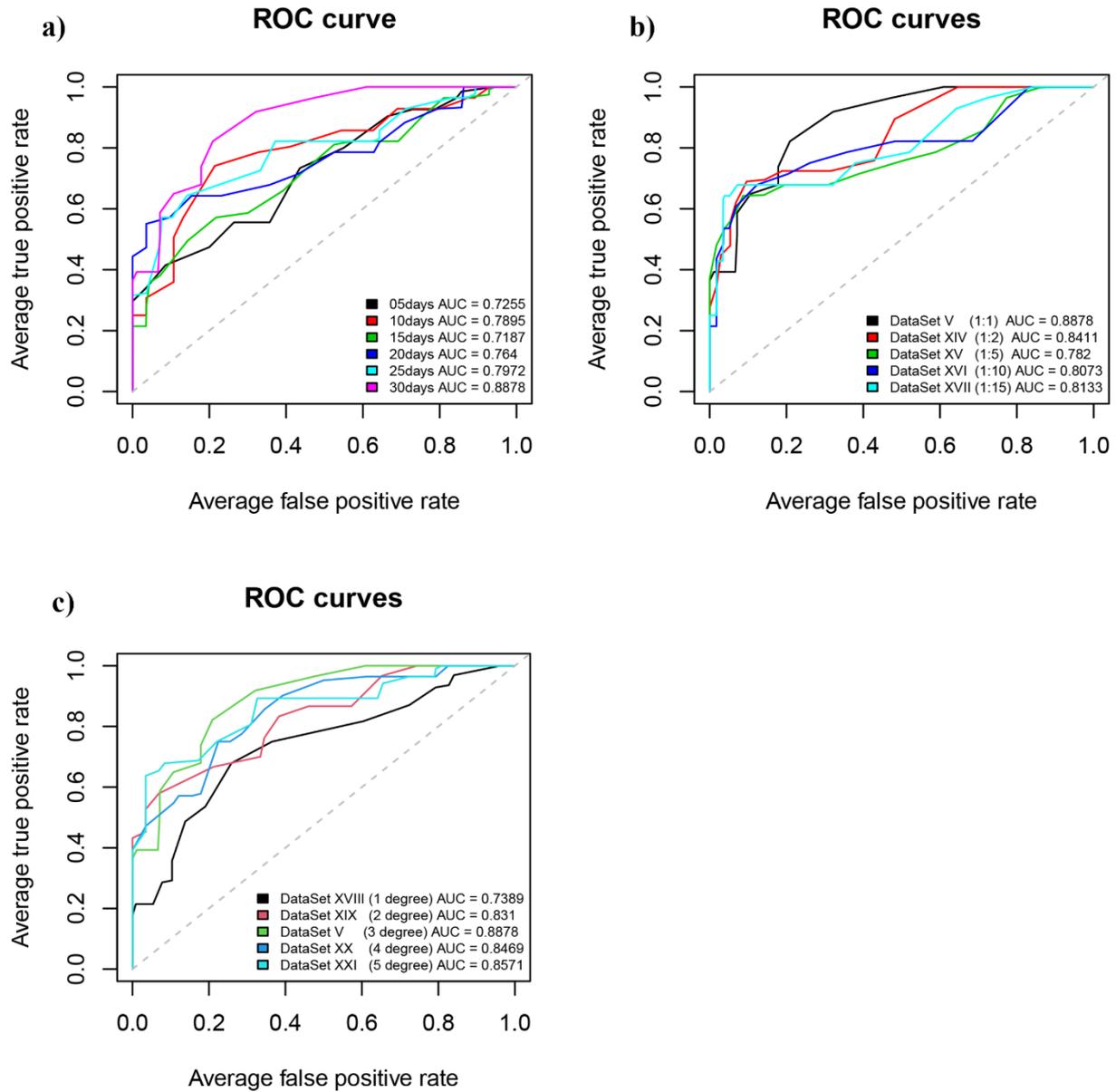

Figure S5 The ROC curves on a) the six datasets with different temporal windows, b) the five unbalanced datasets with different positive to negative ratios, and c) the five datasets of different spatial windows.

Table S1 Satellite's multi-source and multi-parameters information can be utilized in the pre-earthquake anomaly detection.

| Multi-parameters | Data source | temporal coverage | Temporal resolution | spatial coverage | Spatial resolution |
|---|---|---|---|---|---|
| Infrared brightness temperature | FY2-VISSR | Since 2005 | Half an hour | 60°N-60°S, 45°E-165°E （FY2C/2E） 45°E-165°E (FY2D) | 0.1° x 0.1° |
| | NOAA-AVHRR | Since 1994 | Twice a day | global coverage | 1.1km |
| | AUAQ/TERRA-MODIS | Since 2000 | Twice a day | global coverage | 1.0 km |
| Outgoing longwave radiation | NOAA-AVHRR | Since 1994 | Twice a day | global coverage | 1°x 1° 2.5°x 2.5° |
| | FY2-VISSR | Since 2005 | Half an hour | 60°N-60°S, 45°E-165°E (FY2C/2E) 45°E-165°E (FY2D) | 0.1°x 0.1° |
| Surface Temperature | AQUA- AIRS | Since 2002 | Twice a day | global coverage | 1°x 1° |
| | NCEP-NCAR Reanalysis | Since 1948 | Six hours | global coverage | 1.9°x 1.9° |
| | AUAQ/TERRA-MODIS | Since 2000 | Four times a day | global coverage | 1.0 km |
| SLHF | NCEP-NCAR Reanalysis | Since 1948 | Six hours | global coverage | 1.9°x 1.9° |
| Water vapour contents、$O_3$、CO、$CH_4$ | AQUA- AIRS | Since 2002 | Twice a day | global coverage | 1°x 1° |
| $CO_2$ | AQUA- AIRS | Since 2002 | Twice a day | global coverage | 2.5°x 2.5° |

Table S2 Details of ten multi-parameters used in the experiments.

| Vertical Level | Multi-parameters | Physical interpretation | Data obtain | Temporal resolution | Spatial resolution |
|---|---|---|---|---|---|
| Ground surface | MODIS_LST | Land surface temperatures | AQUA- AIRS | Twice a day | 1° x 1° |
| | SurfSkinTemp_AIRS | Surface skin temperature. (Kelvin) | AQUA- AIRS | Twice a day | 1° x 1° |
| Specified height above ground | SurfAirTemp_AIRS | Temperature of the atmosphere at the Earth's surface. (Kelvin) | AQUA- AIRS | Twice a day | 1° x 1° |
| | $H_2O$_AIRS | Water vapour mass mixing ratio at the surface (gm/kg dry air) | AQUA- AIRS | Twice a day | 1° x 1° |
| | $O_3$_AIRS | Total integrated column ozone burden. (Dobson units) | AQUA- AIRS | Twice a day | 1° x 1° |
| | CO_AIRS | Retrieved total column CO. (molecules/$cm^2$) | AQUA- AIRS | Twice a day | 1° x 1° |
| | $CH_4$_AIRS | Retrieved total column CH4. (molecules/$cm^2$) | AQUA- AIRS | Twice a day | 1° x 1° |
| Top of atmosphere | OLR_AIRS | Outgoing longwave radiation flux. (watts/$m^2$) | AQUA- AIRS | Twice a day | 1° x 1° |
| | CLOLR_AIRS | Clear-sky outgoing longwave radiation flux. (watts/$m^2$) | AQUA- AIRS | Twice a day | 1° x 1° |
| | OLR_NOAA | Outgoing longwave radiation flux. (watts/$m^2$) | NOAA-AVHRR | Twice a day | 1° x 1° |

Table S3 Number of label clusters for parameters in every datasets.

| Parameters<br>DataSets | SurfSkinTemp<br>AIRS | SurfAirTemp<br>AIRS | H2O<br>AIRS | O3<br>AIRS | CO<br>AIRS | CH4<br>AIRS | OLR<br>AIRS | CLOLR<br>AIRS | OLR<br>NOAA | MODIS<br>LST |
|---|---|---|---|---|---|---|---|---|---|---|
| **DataSet I** | 11 | 11 | 10 | 10 | 11 | 10 | 11 | 10 | 10 | 4 |
| **DataSet II** | 10 | 10 | 10 | 10 | 11 | 10 | 10 | 10 | 11 | 4 |
| **DataSet III** | 10 | 10 | 10 | 9 | 10 | 10 | 10 | 10 | 9 | 10 |
| **DataSet IV** | 10 | 10 | 10 | 10 | 10 | 10 | 10 | 10 | 10 | 10 |
| **DataSet V** | 10 | 11 | 11 | 19 | 10 | 10 | 11 | 13 | 10 | 4 |
| **DataSet VI** | 11 | 12 | 11 | 19 | 11 | 11 | 11 | 14 | 15 | 4 |
| **DataSet VII** | 9 | 10 | 10 | 10 | 10 | 10 | 10 | 10 | 10 | 10 |
| **DataSet VIII** | 10 | 10 | 10 | 9 | 10 | 10 | 10 | 10 | 10 | 10 |
| Dataset IX | 15 | 11 | 13 | 17 | 17 | 10 | 14 | 14 | 10 | 4 |
| Dataset X | 11 | 10 | 17 | 17 | 17 | 11 | 14 | 17 | 10 | 4 |
| Dataset XI | 10 | 16 | 14 | 10 | 11 | 11 | 11 | 11 | 10 | 5 |
| Dataset XII | 18 | 11 | 10 | 10 | 14 | 10 | 13 | 13 | 11 | 4 |
| Dataset XIII | 12 | 10 | 10 | 15 | 14 | 11 | 10 | 12 | 14 | 4 |
| Dataset XIV | 10 | 15 | 10 | 10 | 10 | 10 | 10 | 15 | 11 | 5 |
| Dataset XV | 10 | 11 | 14 | 13 | 10 | 11 | 14 | 14 | 11 | 5 |
| Dataset XVI | 11 | 16 | 11 | 11 | 11 | 11 | 11 | 11 | 10 | 6 |
| Dataset XVII | 11 | 10 | 13 | 12 | 10 | 11 | 14 | 13 | 11 | 4 |

Table S4 Data sets used as input to the models (DataSet I to DataSet VIII).

| | Spatial Feature | Temporal Feature | Features generation | Earthquakes magnitude | Real earthquakes/ non-seismic events | |
|---|---|---|---|---|---|---|
| **DataSet I** | with its center at the epicenter and a deviation of 3° | 30 days before an earthquake | Time series based features | earthquakes of magnitude 7 or greater | 137 earthquakes/ 137 earthquakes | with aftershocks |
| **DataSet II** | with its center at the epicenter and a deviation of 3° | 30 days before an earthquake | Time series based features | earthquakes of magnitude between 6 and 7 | 1234 earthquakes/ 1234 earthquakes | with aftershocks |
| **DataSet III** | with its center at the epicenter and a deviation of 3° | 30 days before an earthquake | Standard features | earthquakes of magnitude 7 or greater | 137 earthquakes/ 137 earthquakes | with aftershocks |
| **DataSet IV** | with its center at the epicenter and a deviation of 3° | 30 days before an earthquake | Standard features | earthquakes of magnitude between 6 and 7 | 1234 earthquakes/ 1234 earthquakes | with aftershocks |
| **DataSet V** | with its center at the epicenter and a deviation of 3° | 30 days before an earthquake | Time series based features | earthquakes of magnitude 7 or greater | 121 earthquakes/ 121 earthquakes | with aftershocks dropped |
| **DataSet VI** | with its center at the epicenter and a deviation of 3° | 30 days before an earthquake | Time series based features | earthquakes of magnitude between 6 and 7 | 860 earthquakes/ 860 earthquakes | with aftershocks dropped |
| **DataSet VII** | with its center at the epicenter and a deviation of 3° | 30 days before an earthquake | Standard features | earthquakes of magnitude 7 or greater | 121 earthquakes/ 121 earthquakes | with aftershocks dropped |
| **DataSet VIII** | with its center at the epicenter and a deviation of 3° | 30 days before an earthquake | Standard features | earthquakes of magnitude between 6 and 7 | 860 earthquakes/ 860 earthquakes | with aftershocks dropped |

Table S5 Forecasting performance with the six best-benchmarking methods on DataSet I and V (Satellite data of earthquakes of magnitude 7 or greater) with the generated features. Bold: The bold refers to the first place result of all the methods in the comparisons.

| | DataSet I | | | | | | | | | | |
|---|---|---|---|---|---|---|---|---|---|---|---|
| | MCC | R score | AUC | Specificity | Sensitivity | Accuracy | Precision | TN | TP | FN | FP |
| **IBPT** | **0.6581** | **0.6429** | **0.8718** | **0.7143** | **0.9286** | **0.8214** | **0.7647** | **20** | **26** | **2** | **8** |
| RF | 0.5388 | 0.5357 | 0.8316 | 0.7143 | 0.8214 | 0.7679 | 0.7419 | 20 | 23 | 5 | 8 |
| CNN | 0.5657 | 0.5357 | 0.7449 | 0.6071 | 0.9286 | 0.7679 | 0.7027 | 17 | 26 | 2 | 11 |
| FI | 0.5533 | 0.5357 | 0.7997 | 0.8929 | 0.6429 | 0.7679 | 0.8571 | 25 | 18 | 10 | 3 |
| GBM | 0.5013 | 0.5 | 0.7857 | 0.7857 | 0.7143 | 0.75 | 0.7692 | 22 | 20 | 8 | 6 |
| MLP | 0.4903 | 0.4643 | 0.5829 | 0.5714 | 0.8929 | 0.7321 | 0.6757 | 16 | 25 | 3 | 12 |
| | DataSet V | | | | | | | | | | |
| | MCC | R score | AUC | Specificity | Sensitivity | Accuracy | Precision | TN | TP | FN | FP |
| **IBPT** | **0.6429** | **0.6429** | **0.8878** | **0.8214** | **0.8214** | **0.8214** | **0.8214** | **23** | **23** | **5** | **5** |
| RF | 0.6075 | 0.6071 | 0.8642 | 0.8214 | 0.7857 | 0.8036 | 0.8148 | 23 | 22 | 6 | 5 |
| CNN | 0.5 | 0.5 | 0.7895 | 0.75 | 0.75 | 0.75 | 0.75 | 21 | 21 | 7 | 7 |
| FI | 0.3951 | 0.3929 | 0.72 | 0.6429 | 0.75 | 0.6964 | 0.6774 | 18 | 21 | 7 | 10 |
| GBM | 0.6107 | 0.6071 | 0.8272 | 0.75 | 0.8571 | 0.8036 | 0.7742 | 21 | 24 | 4 | 7 |
| MLP | 0.2582 | 0.25 | 0.5893 | 0.5 | 0.75 | 0.625 | 0.6 | 14 | 21 | 7 | 14 |

Table S6 Forecasting performance with the six best-benchmarking methods on DataSet II and VI (Satellite data of earthquakes of magnitude between 6 and 7) with the generated features. Bold: The bold indicates the first place result of all the methods in the comparisons.

| | DataSet II | | | | | | | | | | |
|---|---|---|---|---|---|---|---|---|---|---|---|
| | MCC | R score | AUC | Specificity | Sensitivity | Accuracy | Precision | TN | TP | FN | FP |
| **IBPT** | **0.5958** | **0.5942** | **0.8683** | **0.834** | **0.7602** | **0.7972** | **0.8202** | **206** | **187** | **59** | **41** |
| RF | 0.57 | 0.57 | 0.8664 | 0.7895 | 0.7805 | 0.785 | 0.7869 | 195 | 192 | 54 | 52 |
| CNN | 0.3866 | 0.3797 | 0.7365 | 0.5951 | 0.7846 | 0.6897 | 0.6587 | 147 | 193 | 53 | 100 |
| FI | 0.3943 | 0.3912 | 0.7166 | 0.7571 | 0.6341 | 0.6957 | 0.7222 | 187 | 156 | 90 | 60 |
| GBM | 0.4727 | 0.4726 | 0.7977 | 0.749 | 0.7236 | 0.7363 | 0.7417 | 185 | 178 | 68 | 62 |
| MLP | 0.3064 | 0.3063 | 0.7114 | 0.6437 | 0.6626 | 0.6531 | 0.6494 | 159 | 163 | 83 | 88 |
| | DataSet VI | | | | | | | | | | |
| | MCC | R score | AUC | Specificity | Sensitivity | Accuracy | Precision | TN | TP | FN | FP |
| **IBPT** | **0.5258** | **0.5058** | **0.8211** | **0.8895** | **0.6163** | **0.7529** | **0.848** | **153** | **106** | **66** | **19** |
| RF | 0.4844 | 0.4651 | 0.8136 | 0.8721 | 0.593 | 0.7326 | 0.8226 | 150 | 102 | 70 | 22 |
| CNN | 0.4746 | 0.4244 | 0.7579 | 0.936 | 0.4884 | 0.7122 | 0.8842 | 161 | 84 | 88 | 11 |
| FI | 0.3743 | 0.3547 | 0.7172 | 0.8372 | 0.5174 | 0.6773 | 0.7607 | 144 | 89 | 83 | 28 |
| GBM | 0.4447 | 0.3895 | 0.7625 | 0.936 | 0.4535 | 0.6948 | 0.8764 | 161 | 78 | 94 | 11 |
| MLP | 0.3544 | 0.3081 | 0.6885 | 0.407 | 0.9012 | 0.6541 | 0.6031 | 70 | 155 | 17 | 102 |

Table S7. Forecasting performance with the six best-benchmarking methods on DataSet III and VII (the satellite data of earthquakes of magnitude 7 or greater with the standard features). The bold indicates the first place result of all the methods in the comparisons.

| | DataSet III | | | | | | | | | | |
|---|---|---|---|---|---|---|---|---|---|---|---|
| | MCC | R score | AUC | Specificity | Sensitivity | Accuracy | Precision | TN | TP | FN | FP |
| **IBPT** | **0.5587** | **0.5495** | **0.8104** | **0.6923** | **0.8571** | **0.7778** | **0.75** | **18** | **24** | **4** | **8** |
| RF | 0.593 | 0.5934 | 0.7816 | 0.8077 | 0.7857 | 0.7963 | 0.8148 | 21 | 22 | 6 | 5 |
| CNN | 0.3602 | 0.3159 | 0.6415 | 0.4231 | 0.8929 | 0.6667 | 0.625 | 11 | 25 | 3 | 15 |
| FI | 0.2946 | 0.294 | 0.6078 | 0.6154 | 0.6786 | 0.6481 | 0.6552 | 16 | 19 | 9 | 10 |
| GBM | 0.5189 | 0.5192 | 0.7679 | 0.7692 | 0.75 | 0.7593 | 0.7778 | 20 | 21 | 7 | 6 |
| MLP | 0.3602 | 0.3159 | 0.6291 | 0.4231 | 0.8929 | 0.6667 | 0.625 | 11 | 25 | 3 | 15 |
| | DataSet VII | | | | | | | | | | |
| | MCC | R score | AUC | Specificity | Sensitivity | Accuracy | Precision | TN | TP | FN | FP |
| **IBPT** | **0.5754** | **0.5493** | **0.8227** | **0.9286** | **0.6207** | **0.7719** | **0.9** | **26** | **18** | **11** | **2** |
| RF | 0.4067 | 0.4052 | 0.7278 | 0.75 | 0.6552 | 0.7018 | 0.7308 | 21 | 19 | 10 | 7 |
| CNN | 0.4442 | 0.4409 | 0.7069 | 0.7857 | 0.6552 | 0.7193 | 0.76 | 22 | 19 | 10 | 6 |
| FI | 0.268 | 0.2586 | 0.6552 | 0.5 | 0.7586 | 0.6316 | 0.6111 | 14 | 22 | 7 | 14 |
| GBM | 0.4414 | 0.436 | 0.7457 | 0.6429 | 0.7931 | 0.7193 | 0.697 | 18 | 23 | 6 | 10 |
| MLP | 0.3784 | 0.3372 | 0.6715 | 0.4419 | 0.8953 | 0.6686 | 0.616 | 76 | 154 | 18 | 96 |

Table S8 Forecasting performance with the six best-benchmarking methods on DataSet IV and VIII (the satellite data of earthquakes of magnitude between 6 and 7 with the standard features). The bold refers to the first place result of all the methods in the comparisons.

| | MCC | R score | AUC | Specificity | Sensitivity | Accuracy | Precision | TN | TP | FN | FP |
|---|---|---|---|---|---|---|---|---|---|---|---|
| | **DataSet IV** | | | | | | | | | | |
| **IBPT** | **0.4813** | **0.4808** | **0.8156** | **0.7166** | **0.7642** | **0.7404** | **0.7287** | **177** | **188** | **58** | **70** |
| RF | 0.4039 | 0.392 | 0.7408 | 0.5749 | 0.8171 | 0.6957 | 0.6569 | 142 | 201 | 45 | 105 |
| CNN | 0.4457 | 0.4405 | 0.7342 | 0.6437 | 0.7967 | 0.7201 | 0.6901 | 159 | 196 | 50 | 88 |
| FI | 0.2536 | 0.2535 | 0.6492 | 0.6356 | 0.6179 | 0.6268 | 0.6281 | 157 | 152 | 94 | 90 |
| GBM | 0.4816 | 0.4808 | 0.7917 | 0.7126 | 0.7683 | 0.7404 | 0.7269 | 176 | 189 | 57 | 71 |
| MLP | 0.3273 | 0.3189 | 0.6652 | 0.5466 | 0.7724 | 0.6592 | 0.6291 | 135 | 190 | 56 | 112 |
| | **DataSet VIII** | | | | | | | | | | |
| | MCC | R score | AUC | Specificity | Sensitivity | Accuracy | Precision | TN | TP | FN | FP |
| **IBPT** | **0.4905** | **0.4884** | **0.8143** | **0.7907** | **0.6977** | **0.7442** | **0.7692** | **136** | **120** | **52** | **36** |
| RF | 0.402 | 0.3547 | 0.626 | 0.4419 | 0.9128 | 0.6773 | 0.6206 | 76 | 157 | 15 | 96 |
| CNN | 0.4304 | 0.407 | 0.7136 | 0.5407 | 0.8663 | 0.7035 | 0.6535 | 93 | 149 | 23 | 79 |
| FI | 0.228 | 0.2151 | 0.6277 | 0.4419 | 0.7733 | 0.6076 | 0.5808 | 76 | 133 | 39 | 96 |
| GBM | 0.4533 | 0.436 | 0.7921 | 0.8547 | 0.5814 | 0.718 | 0.8 | 147 | 100 | 72 | 25 |
| MLP | 0.3306 | 0.3256 | 0.7294 | 0.75 | 0.5756 | 0.6628 | 0.6972 | 129 | 99 | 73 | 43 |

Table S9 Forecasting performance with the benchmarked eight datasets using IBPT.

| | DataSets | MCC | R score | AUC | Specificity | Sensitivity | Accuracy | Precision | TN | TP | FN | FP |
|---|---|---|---|---|---|---|---|---|---|---|---|---|
| Proposed Feature | DataSet I | 0.6581 | 0.6429 | 0.8718 | 0.7143 | 0.9286 | 0.8214 | 0.7647 | 20 | 26 | 2 | 8 |
| | DataSet II | 0.5958 | 0.5942 | 0.8683 | 0.834 | 0.7602 | 0.7972 | 0.8202 | 206 | 187 | 59 | 41 |
| | DataSet V | 0.6429 | 0.6429 | 0.8878 | 0.8214 | 0.8214 | 0.8214 | 0.8214 | 23 | 23 | 5 | 5 |
| | DataSet VI | 0.5258 | 0.5058 | 0.8211 | 0.8895 | 0.6163 | 0.7529 | 0.848 | 153 | 106 | 66 | 19 |
| Standard Feature | DataSet III | 0.5587 | 0.5495 | 0.8104 | 0.6923 | 0.8571 | 0.7778 | 0.75 | 18 | 24 | 4 | 8 |
| | DataSet IV | 0.4813 | 0.4808 | 0.8156 | 0.7166 | 0.7642 | 0.7404 | 0.7287 | 177 | 188 | 58 | 70 |
| | DataSet VII | 0.5754 | 0.5493 | 0.8227 | 0.9286 | 0.6207 | 0.7719 | 0.9 | 26 | 18 | 11 | 2 |
| | DataSet VIII | 0.4905 | 0.4884 | 0.8143 | 0.7907 | 0.6977 | 0.7442 | 0.7692 | 136 | 120 | 52 | 36 |

Table S10 Forecasting performance with the six best-benchmarking methods on DataSet I and I-nonoverlap (Satellite data of earthquakes of magnitude 7 or greater) with the generated features. Bold: The bold indicates the first place result of all the methods in the comparisons.

| | MCC | R score | AUC | Specificity | Sensitivity | Accuracy | Precision | TN | TP | FN | FP |
|---|---|---|---|---|---|---|---|---|---|---|---|
| DataSet I | | | | | | | | | | | |
| | MCC | R score | AUC | Specificity | Sensitivity | Accuracy | Precision | TN | TP | FN | FP |
| **IBPT** | **0.6581** | **0.6429** | **0.8718** | **0.7143** | **0.9286** | **0.8214** | **0.7647** | **20** | **26** | **2** | **8** |
| RF | 0.5388 | 0.5357 | 0.8316 | 0.7143 | 0.8214 | 0.7679 | 0.7419 | 20 | 23 | 5 | 8 |
| CNN | 0.5657 | 0.5357 | 0.7449 | 0.6071 | 0.9286 | 0.7679 | 0.7027 | 17 | 26 | 2 | 11 |
| FI | 0.5533 | 0.5357 | 0.7997 | 0.8929 | 0.6429 | 0.7679 | 0.8571 | 25 | 18 | 10 | 3 |
| GBM | 0.5013 | 0.5 | 0.7857 | 0.7857 | 0.7143 | 0.75 | 0.7692 | 22 | 20 | 8 | 6 |
| MLP | 0.4903 | 0.4643 | 0.5829 | 0.5714 | 0.8929 | 0.7321 | 0.6757 | 16 | 25 | 3 | 12 |
| DataSet I-nonoverlap | | | | | | | | | | | |
| | MCC | R score | AUC | Specificity | Sensitivity | Accuracy | Precision | TN | TP | FN | FP |
| **IBPT** | **0.6075** | **0.6071** | **0.8170** | **0.8214** | **0.7857** | **0.8036** | **0.8148** | **23** | **22** | **6** | **5** |
| RF | 0.5197 | 0.463 | 0.7606 | 0.5 | 0.963 | 0.7273 | 0.65 | 14 | 26 | 1 | 14 |
| CNN | 0.3299 | 0.2857 | 0.6932 | 0.3929 | 0.8929 | 0.6429 | 0.5952 | 11 | 25 | 3 | 17 |
| FI | 0.4588 | 0.4286 | 0.7143 | 0.8929 | 0.5357 | 0.7143 | 0.8333 | 25 | 15 | 13 | 3 |
| GBM | 0.5678 | 0.5185 | 0.7634 | 0.963 | 0.5556 | 0.7593 | 0.9375 | 26 | 15 | 12 | 1 |
| MLP | 0.3951 | 0.3929 | 0.7398 | 0.6429 | 0.75 | 0.6964 | 0.6774 | 18 | 21 | 7 | 10 |

Table S11 Forecasting performance with the six best-benchmarking methods on DataSet II and II-nonoverlap (Satellite data of earthquakes of magnitude between 6 and 7) with the generated features. Bold: The bold indicates the first place result of all the methods in the comparisons.

| | DataSet II | | | | | | | | | | |
|---|---|---|---|---|---|---|---|---|---|---|---|
| | MCC | R score | AUC | Specificity | Sensitivity | Accuracy | Precision | TN | TP | FN | FP |
| **IBPT** | **0.5958** | **0.5942** | **0.8683** | **0.834** | **0.7602** | **0.7972** | **0.8202** | **206** | **187** | **59** | **41** |
| RF | 0.57 | 0.57 | 0.8664 | 0.7895 | 0.7805 | 0.785 | 0.7869 | 195 | 192 | 54 | 52 |
| CNN | 0.3866 | 0.3797 | 0.7365 | 0.5951 | 0.7846 | 0.6897 | 0.6587 | 147 | 193 | 53 | 100 |
| FI | 0.3943 | 0.3912 | 0.7166 | 0.7571 | 0.6341 | 0.6957 | 0.7222 | 187 | 156 | 90 | 60 |
| GBM | 0.4727 | 0.4726 | 0.7977 | 0.749 | 0.7236 | 0.7363 | 0.7417 | 185 | 178 | 68 | 62 |
| MLP | 0.3064 | 0.3063 | 0.7114 | 0.6437 | 0.6626 | 0.6531 | 0.6494 | 159 | 163 | 83 | 88 |
| | DataSet II –nonoverlap | | | | | | | | | | |
| | MCC | R score | AUC | Specificity | Sensitivity | Accuracy | Precision | TN | TP | FN | FP |
| **IBPT** | **0.6176** | **0.6000** | **0.8613** | **0.9184** | **0.6816** | **0.8000** | **0.8930** | **225** | **167** | **78** | **20** |
| RF | 0.5839 | 0.5673 | 0.8591 | 0.9020 | 0.6653 | 0.7837 | 0.8717 | 221 | 163 | 82 | 24 |
| CNN | 0.4287 | 0.4082 | 0.7651 | 0.8571 | 0.551 | 0.7041 | 0.7941 | 210 | 135 | 110 | 35 |
| FI | 0.2451 | 0.2449 | 0.6532 | 0.6041 | 0.6408 | 0.6224 | 0.6181 | 148 | 157 | 88 | 97 |
| GBM | 0.4261 | 0.4122 | 0.7721 | 0.8327 | 0.5796 | 0.7061 | 0.7760 | 204 | 142 | 103 | 41 |
| MLP | 0.2816 | 0.2122 | 0.5669 | 0.2776 | 0.9347 | 0.6061 | 0.5640 | 68 | 229 | 16 | 177 |

Table S12 Data sets used as input to the models (Dataset IX to Dataset XXI).

| | Spatial Feature | Temporal Feature | Features generation | Earthquakes magnitude | Number of real earthquakes/ artificial non-seismic events |
|---|---|---|---|---|---|
| Dataset IX | with a deviation of 3° | 25 days before an earthquake | Time series based features | earthquakes of magnitude 7 or greater | 121 earthquakes/ 121 earthquakes |
| Dataset X | with a deviation of 3° | 20 days before an earthquake | Time series based features | earthquakes of magnitude 7 or greater | 121 earthquakes/ 121 earthquakes |
| Dataset XI | with a deviation of 3° | 15 days before an earthquake | Time series based features | earthquakes of magnitude 7 or greater | 121 earthquakes/ 121 earthquakes |
| Dataset XII | with a deviation of 3° | 10 days before an earthquake | Time series based features | earthquakes of magnitude 7 or greater | 121 earthquakes/ 121 earthquakes |
| Dataset XIII | with a deviation of 3° | 05 days before an earthquake | Time series based features | earthquakes of magnitude 7 or greater | 121 earthquakes/ 121 earthquakes |
| Dataset XIV | with a deviation of 3° | 30 days before an earthquake | Time series based features | earthquakes of magnitude 7 or greater | 121 earthquakes/ 242 earthquakes |
| Dataset XV | with a deviation of 3° | 30 days before an earthquake | Time series based features | earthquakes of magnitude 7 or greater | 121 earthquakes/ 605 earthquakes |
| Dataset XVI | with a deviation of 3° | 30 days before an earthquake | Time series based features | earthquakes of magnitude 7 or greater | 121 earthquakes/ 1210 earthquakes |
| Dataset XVII | with a deviation of 3° | 30 days before an earthquake | Time series based features | earthquakes of magnitude 7 or greater | 121 earthquakes/ 1815 earthquakes |
| DataSet XVIII | with a deviation of 1° | 30 days before an earthquake | Time series based features | earthquakes of magnitude 7 or greater | 121 earthquakes/ 121 earthquakes |
| DataSet XIX | with a deviation of 2° | 30 days before an earthquake | Time series based features | earthquakes of magnitude 7 or greater | 121 earthquakes/ 121 earthquakes |
| DataSet XX | with a deviation of 4° | 30 days before an earthquake | Time series based features | earthquakes of magnitude 7 or greater | 121 earthquakes/ 121 earthquakes |
| DataSet XXI | with a deviation of 5° | 30 days before an earthquake | Time series based features | earthquakes of magnitude 7 or greater | 121 earthquakes/ 121 earthquakes |

Table S13. Forecasting performance with the six datasets of different temporal windows using IBPT.

| DataSets | Temporal window | MCC | R score | AUC | Specificity | Sensitivity | Accuracy | Precision | TN | TP | FN | FP |
|---|---|---|---|---|---|---|---|---|---|---|---|---|
| DataSet V | 30 days | 0.6429 | 0.6429 | 0.8878 | 0.8214 | 0.8214 | 0.8214 | 0.8214 | 23 | 23 | 5 | 5 |
| Dataset IX | 25 days | 0.5533 | 0.5357 | 0.7972 | 0.8929 | 0.6429 | 0.7679 | 0.8571 | 25 | 18 | 10 | 3 |
| Dataset X | 20 days | 0.5826 | 0.5357 | 0.764 | 0.9643 | 0.5714 | 0.7679 | 0.9412 | 27 | 16 | 12 | 1 |
| Dataset XI | 15 days | 0.3824 | 0.3571 | 0.7188 | 0.8571 | 0.5 | 0.6786 | 0.7778 | 24 | 14 | 14 | 4 |
| Dataset XII | 10 days | 0.5361 | 0.5357 | 0.7895 | 0.7857 | 0.75 | 0.7679 | 0.7778 | 22 | 21 | 7 | 6 |
| Dataset XIII | 05 days | 0.3953 | 0.336 | 0.7255 | 0.9286 | 0.4074 | 0.6727 | 0.8462 | 26 | 11 | 16 | 2 |

Table S14. Forecasting performance with the five datasets of different spatial windows using IBPT.

| DataSets | Spatial window | MCC | R score | AUC | Specificity | Sensitivity | Accuracy | Precision | TN | TP | FN |
|---|---|---|---|---|---|---|---|---|---|---|---|
| DataSet XVIII | 1 degree | 0.4388 | 0.4372 | 0.7389 | 0.7586 | 0.6786 | 0.7193 | 0.7308 | 22 | 19 | 9 |
| DataSet XIX | 2 degree | 0.5893 | 0.5644 | 0.831 | 0.931 | 0.6333 | 0.7797 | 0.9048 | 27 | 19 | 11 |
| DataSet V | 3 degree | 0.6429 | 0.6429 | 0.8878 | 0.8214 | 0.8214 | 0.8214 | 0.8214 | 23 | 23 | 5 |
| DataSet XX | 4 degree | 0.5533 | 0.5357 | 0.8469 | 0.6429 | 0.8929 | 0.7679 | 0.7143 | 18 | 25 | 3 |
| DataSet XXI | 5 degree | 0.6318 | 0.6096 | 0.8571 | 0.931 | 0.6786 | 0.807 | 0.9048 | 27 | 19 | 9 |

Table S15. Forecasting performance with the five datasets of different rate of true earthquakes and artificial-non earthquakes using IBPT.

| DataSets | Positive to Negative ratio | MCC | R score | AUC | Specificity | Sensitivity | Accuracy | Precision | TN | TP | FN |
|---|---|---|---|---|---|---|---|---|---|---|---|
| DataSet V | 1:1 | 0.6429 | 0.6429 | 0.8878 | 0.8214 | 0.8214 | 0.8214 | 0.8214 | 23 | 23 | 5 |
| Dataset XIV | 1:2 | 0.6247 | 0.6004 | 0.8411 | 0.9107 | 0.6897 | 0.8353 | 0.8 | 51 | 20 | 9 |
| Dataset XV | 1:5 | 0.6145 | 0.5727 | 0.782 | 0.9298 | 0.6429 | 0.8353 | 0.8182 | 53 | 18 | 10 |
| Dataset XVI | 1:10 | 0.6169 | 0.5909 | 0.8073 | 0.9123 | 0.6786 | 0.8353 | 0.7917 | 52 | 19 | 9 |
| Dataset XVII | 1:15 | 0.6717 | 0.6259 | 0.8133 | 0.9474 | 0.6786 | 0.8588 | 0.8636 | 54 | 19 | 9 |

Table S16. Previous studies using machine learning for earthquake prediction from the satellite data. BPNN: back-propagation neural network. MARBDP: mining association rule based on dynamic pruning. DEMETER: Detection of Electro-Magnetic Emissions Transmitted from Earthquake Regions. CSES: China Seismo-Electromagnetic Satellite. GNSS: Global Navigation Satellite System. TEC: Total electron content. AIRS: Atmospheric Infrared Sounder. NOAA: National Oceanic and Atmospheric Administration.

| Study | Study Area | Study Period | Satellite Data | Model | Objective and Performance |
|---|---|---|---|---|---|
| Xu et al. (2010) | The globe | 2007–2008 | DEMETER | BPNN | Predict seismic events in 2008. Accuracy: 69.96%. |
| Wang et al. (2014) | Taiwan, China | January 2008–June 2008 | DEMETER | MARBDP | Predict earthquakes of Ms >5.0 from January 2008 to June 2008. sensitivity: 70.01% |
| Li et al. (2020) | The globe | June 2004 to December 2010. | DEMETER and CSES | statistical method | Electromagnetic pre-earthquake perturbations detection. False positive rate: 50.2% |
| Ouyang et al. (2020) | The globe | May 2005 to November 2010 | DEMETER | Superposed epoch analysis | Electromagnetic pre-earthquake perturbations detection. Accuracy: 34% |
| Liu et al. (2000) | Taiwan | 1994 to 1999 | GNSS TEC | the interquartile range (IQR) | TEC pre-earthquake anomalies detection. Accuracy: 73.8% |
| Our proposed method | The globe | January 2006 to December 2013 | NASA AIRS and NOAA | IBPT | Earthquakes prediction. Accuracy: 82.14% Sensitivity: 92.86% False positive rate: 28.57% |